\documentclass[aps,
prb,
reprint,
preprintnumbers,
superscriptaddress,
amsfonts,
amssymb,
amsmath
]{revtex4-2}

\usepackage{graphicx}
\usepackage{dcolumn}
\usepackage{bm}

\usepackage[utf8]{inputenc}
\usepackage[T1]{fontenc}
\usepackage{mathptmx}
\usepackage{etoolbox}

\usepackage{bm,latexsym,mathrsfs,enumerate,bigints,eucal}
\usepackage{upgreek}
\usepackage{float}
\usepackage[table,x11names]{xcolor}
\usepackage[breaklinks=true,
unicode=true,
urlcolor = RoyalBlue4,
colorlinks = true,
citecolor = RoyalBlue4,
linkcolor = blue
]{hyperref}
\usepackage{graphicx}
\usepackage{chemformula}
\usepackage{siunitx}
\usepackage[page]{appendix} 
\usepackage{mathtools}
\usepackage[frak=boondox,scr=boondox]{mathalfa} 
\usepackage{savesym}
\usepackage[most]{tcolorbox}
\savesymbol{comment}

\usepackage{xcolor, soul}
\sethlcolor{orange!30}

\DeclareMathOperator{\am}{am}
\DeclareMathOperator{\sn}{sn}
\DeclareMathOperator{\cn}{cn}

\DeclareMathOperator{\dn}{dn}

\DeclareMathOperator{\ellipticK}{K}
\DeclareMathOperator{\ellipticE}{E}
\renewcommand{\vec}[1]{\pmb{#1}}
\begin{document}

\definecolor{cgreen}{rgb}{0.11, 0.35, 0.2}
\definecolor{blue0804}{HTML}{16561b}
\definecolor{gr2c}{HTML}{748f10}
\definecolor{br1404}{HTML}{743a0b}

\title{Field-induced spin reorientation transitions in antiferromagnetic ring-shaped spin chains}

\author{Yelyzaveta~A.~Borysenko}
\affiliation{Taras Shevchenko National University of Kyiv, 01601 Kyiv, Ukraine}
\affiliation{Helmholtz-Zentrum Dresden-Rossendorf e.V., Institute of Ion Beam Physics and Materials Research, 01328 Dresden, Germany}
\affiliation{Institute for Theoretical Solid State Physics, Leibniz IFW Dresden, 01069 Dresden, Germany}

\author{Denis~D.~Sheka}
\affiliation{Taras Shevchenko National University of Kyiv, 01601 Kyiv, Ukraine}

\author{J\"{u}rgen~Fassbender}
\affiliation{Helmholtz-Zentrum Dresden-Rossendorf e.V., Institute of Ion Beam Physics and Materials Research, 01328 Dresden, Germany}

\author{Jeroen van den Brink}
\affiliation{Institute for Theoretical Solid State Physics, Leibniz IFW Dresden, 01069 Dresden, Germany}
\affiliation{Institute for Theoretical Physics and W\"{u}rzburg-Dresden Cluster of Excellence ct.qmat, Technische Universit\"{a}t Dresden, 01069 Dresden, Germany}

\author{Denys~Makarov}
\affiliation{Helmholtz-Zentrum Dresden-Rossendorf e.V., Institute of Ion Beam Physics and Materials Research, 01328 Dresden, Germany}

\author{Oleksandr~V.~Pylypovskyi}
\email{o.pylypovskyi@hzdr.de}
\affiliation{Helmholtz-Zentrum Dresden-Rossendorf e.V., Institute of Ion Beam Physics and Materials Research, 01328 Dresden, Germany}
\affiliation{Kyiv Academic University, 03142 Kyiv, Ukraine}

\date{August 4, 2022}
	
\begin{abstract} 
	Easy axis antiferromagnets are robust against external magnetic fields of moderate strength. Spin reorientations in strong fields can provide an insight into more subtle properties of antiferromagnetic materials, which are often hidden by their high ground state symmetry. Here, we investigate theoretically effects of curvature in ring-shaped antiferromagnetic achiral anisotropic spin chains in strong magnetic fields. We identify the geometry-governed
	helimagnetic phase transition above the spin-flop field between vortex and onion states. The curvature-induced Dzyaloshinskii--Moriya interaction results in the spin-flop transition being of first- or second-order depending on the ring curvature. Spatial inhomogeneity of the N\'{e}el vector in the spin-flop phase generates weak ferromagnetic response in the plane perpendicular to the applied magnetic field. Our work contributes to the understanding of the physics of curvilinear antiferromagnets in magnetic fields and guides prospective experimental studies of geometrical effects relying on spin-chain-based nanomagnets.
\end{abstract}

\maketitle
\section{Introduction}

Antiferromagnets (AFMs) represent a broad class of multisublattice magnetic materials, whose magnetic symmetry group contains an element of sublattice permutation~\cite{Turov65, Ivanov05a, Smejkal22}.
In addition to their application potential for high speed and low power magnetic memory and logic devices~\cite{Baltz18,Gomonay17, Smejkal22}, AFMs are complex nonlinear systems, which makes them appealing for fundamental research. This includes studies of material's properties considering crystal symmetries~\cite{Turov01en}, magnetization dynamics and topological solitons~\cite{Kosevich90,Borisov11a}. Being robust against moderate magnetic fields, AFMs possess a family of spin-flop transitions characterized by the reorientation of the N\'{e}el order parameter in sufficiently strong fields. The spin-flop phase is a characteristic feature of the given AFM, which reveals the presence of additional anisotropy axes~\cite{Pisarev97, Woernle21} and can support magnetic solitons~\cite{Kosevich90, Borisov11a,Gomonay18}. In the vicinity of or as a consequence of the phase transition, magnetic responses of AFMs can be modified. Indeed, entering the spin-flop phase reduces the potential barrier for magnetoelectric switching~\cite{Oh14}, enables long-distance spin-transport~\cite{Lebrun18}, enhances the skyrmion lifetime~\cite{Bessarab19} and strengthens the magnetoelectric coupling~\cite{Upadhyay18}. These effects have been intensively studied for bulk AFMs, extended two-dimensional systems and straight spin chains~\cite{Ivanov05a}. In particular, in the case of a straight anisotropic 1D wire with easy axis anisotropy, the spin-flop is the first order transition between different uniform states. Furthermore, there are studies which address quantum effects and cover the influence of sample boundaries and defects on the spin-flop transition~\cite{Peters10, Prokhnenko21}.

The geometry of a magnetic sample provides an additional degree of freedom to tune its anisotropic and chiral responses~\cite{Fischer20,Sheka21b,Streubel21,Makarov22}. Geometrical bends and twists in intrinsically achiral antiferromagnetic spin chains enable helimagnetic responses in the ground states~\cite{Pylypovskyi20}. Geometry-governed modifications of the linear spin dynamics suggest a possibility to form Bose--Einstein condensates for magnons in $k$-space in helix-shaped chains~\cite{Pylypovskyi20} and allow tuning the propagation direction of spin waves of different polarizations~\cite{Wu22}. Being highly sensitive to boundary conditions, AFMs support a variety of non-collinear spin textures in the ring geometry including the M\"{o}bius state~\cite{Castillo-Sepulveda17}. AFM domain walls~\cite{Yershov22a} and skyrmions~\cite{Yershov22} in curvilinear antiferromagnetic thin films are affected by curvature gradients. Beyond the effects predicted for curvilinear antiferromagnets within the $\sigma$-model~\cite{Pylypovskyi20,Yershov22a}, curvilinear AFM spin chains bring about the geometry-governed weak ferromagnetism~\cite{Pylypovskyi21f}. However, spin reorientation transitions in curvilinear AFMs exposed to strong magnetic fields remain unexplored. 

Here, we investigate field-induced spin reorientation transitions in curvilinear ring-shaped intrinsically achiral anisotropic antiferromagnetic spin chains with even number of spins. This includes the highly symmetric case of the field applied along the ring axis and a finite angle between this axis and magnetic field. Using the methodology of curvilinear magnetism, we show that the spin-flop state in a ring geometry enables a helimagnetic transition between the locally homogeneous (vortex) and periodic (onion) AFM textures, which is controlled by the ring curvature. The spin-flop transition for a large enough curvature is supplemented by an intermediate canted state, which we associate with the curvature-induced exchange-driven
Dzyaloshinskii--Moriya interaction (DMI). The description of the curvature-induced weak ferromagnetic response in the spin-flop phase is provided as well. 

\begin{figure*}
\includegraphics[width=\linewidth]{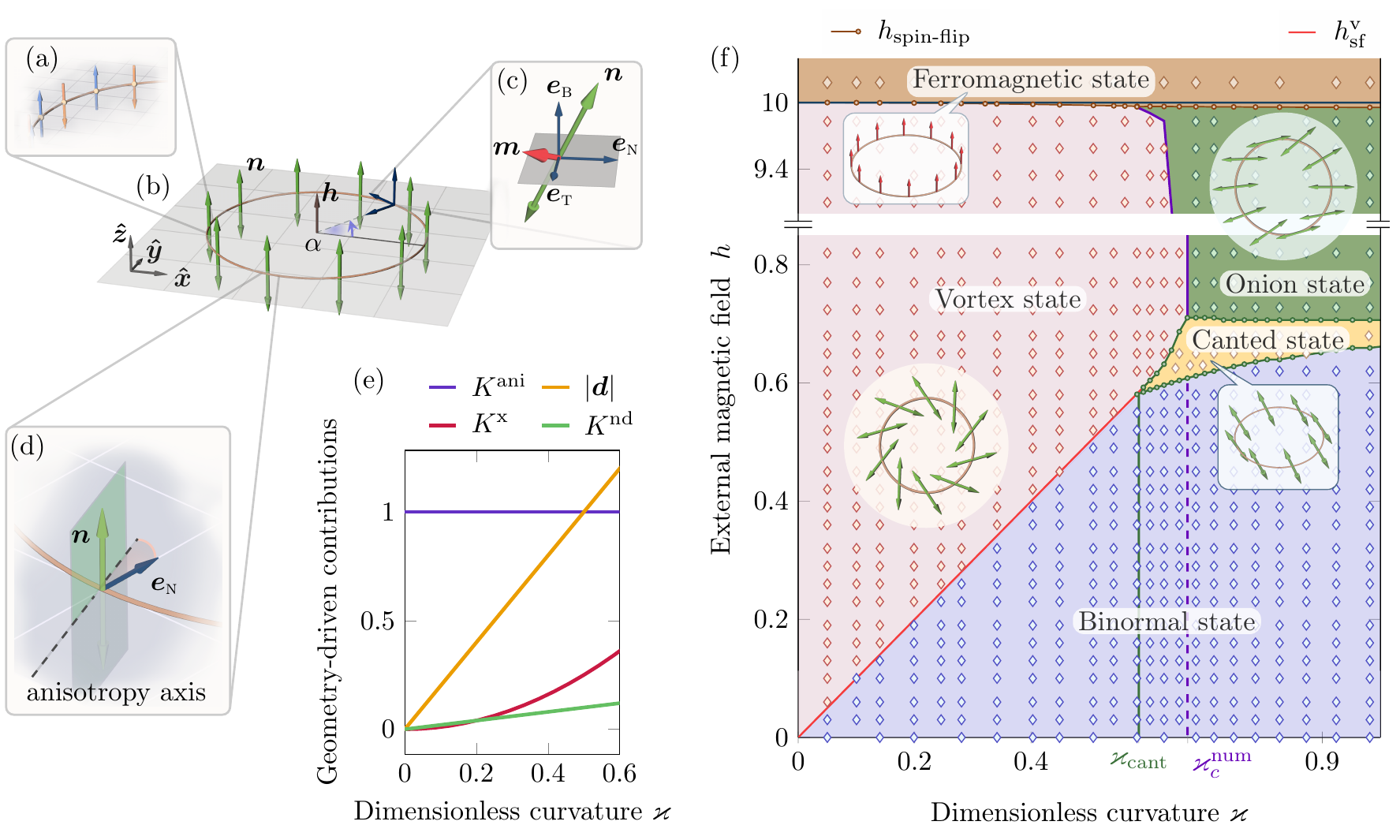}
\caption{\textbf{Antiferromagnetic (AFM) spin chain shaped as a ring exposed to an external magnetic field applied perpendicular to the ring plane.} 
		(a)~Schematics of the discrete model with arrows of blue and orange colors corresponding to the two AFM sublattices.
		(b)~AFM ring in the ground (binormal) state. Double green arrow represents the orientation of the N\'{e}el vector $\vec{n}$. Azimuthal angle $\alpha$ measures coordinate along the ring.
		(c)~TNB and local magnetic reference frames. 
		(d)~Tilt of the anisotropy axes. The green plane indicates the easy plane for the N\'{e}el vector.
		(e)~Relation between the  geometry-governed anisotropic ($K^\text{ani}$, $K^\text{x}$, $K^\text{nd}$) and chiral ($\vec{d}$) contributions in the energy functional. 
		(f)~Equilibrium magnetic textures in the field $h$ applied perpendicular to the ring plane. Symbols represent the data obtained from spin-lattice simulations. Solid black line corresponds to the spin-flip field $h^\text{x} = 1/\varsigma$ and red line is the spin-flop field for small curvatures $h_\text{sf}^\text{v} = \varkappa$. 
	}
	\label{fig:phase_d_hz}
\end{figure*} 

The paper is organized as follows. In Section~\ref{sec:model}, the model of ring-shaped AFM spin chains is introduced. In Section~\ref{sec:perpendicular}, spin-flop phases in the magnetic field applied along the ring axis (the most symmetric case) are discussed. The behavior of the spin chain in the field applied under an angle to the ring axis is considered in Section~\ref{sec:tilted}. Estimations of the characteristic parameters for real nanomagnets are provided in Section~\ref{sec:summary}. In appendices, we summarize further details of spin-lattice simulations (Appendix~\ref{sec:simulations}), offer information on the transition between the discrete and continuum model (Appendix~\ref{sec:discrete}), and describe the spin-flop vortex phase (Appendix~\ref{sec:vortex}).

\section{The model}
\label{sec:model}

We consider an intrinsically achiral ring-shaped spin chain with even number of magnetic moments $N$ and the nearest-neighbor AFM exchange integral $\mathscr{J}$. This leads to the appearance of two ferromagnetically ordered sublattices with opposite directions of magnetic moments, see Fig.~\ref{fig:phase_d_hz}(a). The lattice can be considered as a set of dimers with antiparallel orientation of spins in each dimer. The latter allows to use an alternative representation of the magnetic ordering in terms of the reduced N\'{e}el vector $\vec{n}(\vec{r})$ and reduced magnetization vector $\vec{m}(\vec{r})$ (with $\vec{r}$ being the radius-vector) introduced as the difference and sum of magnetic moments of each AFM dimer along the chain, respectively [Fig.~\ref{fig:phase_d_hz}(b)]. These order parameters can be introduced for a 1D AFM system in such a way to obey the relations $\vec{n}\cdot\vec{m} = 0$ and $\vec{n}^2+\vec{m}^2 = 1$. In the following, we limit our discussion to the case of the single-ion anisotropy with the hard axis of anisotropy along the tangential direction to the ring and spin-lattice coefficient $\mathscr{K}$. The sample's geometry is characterized by the curvature $\kappa = 2\pi/(Na_0)$ with $a_0$ being the distance between the neighboring magnetic moments, represented by spins of length $S$. The local reference frame (referred to as the TNB frame) consists of the tangential $\vec{e}_\textsc{t}$, normal $\vec{e}_\textsc{n}$ and binormal $\vec{e}_\textsc{b}$ vectors, see Fig.~\ref{fig:phase_d_hz}(b) and~\ref{fig:phase_d_hz}(c). The azimuthal angle $\alpha$, which measures the coordinate along the ring, is counted counter-clockwise starting from the $x$-axis [Fig.~\ref{fig:phase_d_hz}(b)].

In the continuum limit of the sufficiently large magnetic length $\ell/a_0 = \sqrt{|\mathscr{J}/\mathscr{K}|} \gg 1$, the magnetic energy of the ring reads
\begin{equation}\label{eq:en-general}
		E/E_0 = \int\left( \mathcal{E}_\text{x}+\mathcal{E}_\text{an}+\mathcal{E}_\text{f} \right) \mathrm{d} \alpha
\end{equation}
with $E_0 = \mathscr{K} S^2/(2 a_0)$, see Appendix~\ref{sec:simulations} for the respective spin-lattice Hamiltonian and details of simulations. The exchange energy density $\mathcal{E}_\text{x} = \vec{m}^2/\varsigma^2 +2(\vec{n}'^2-\vec{m}'^2) + 2\vec{n}' \cdot \vec{m}/\varsigma$ includes the uniform exchange assuring the absence of magnetization in equilibrium, the inhomogeneous exchange responsible for the stabilization of the uniform ground state without external fields, and the lifting term specific for 1D antiferromagnets and responsible for the appearance of the finite magnetization at non-collinear textures~\cite{Pylypovskyi21f, Papanicolaou95}. The prime denotes the spatial derivative $\vec{n}' = \varkappa \partial_{\alpha} \vec{n}$ with $\varkappa = \kappa \ell$ being the dimensionless curvature. The so-called expansion coefficient $\varsigma = a_0/(2\ell)=\sqrt{|\mathscr{K}/(4\mathscr{J})|}$ characterizes the relation between the effective exchange and anisotropy fields. Associating the N\'{e}el vector with the first magnetic moment in each dimer, the intrinsic anisotropy reads $\mathcal{E}_\text{an} = n_\textsc{t}^2 + K^\text{nd} n_\textsc{t}n_\textsc{n}$, where $K^\text{nd} = 2\varsigma\varkappa$~\cite{Pylypovskyi21f}. Here, the hard axis $\vec{e}_\text{ani} = \vec{e}_\textsc{t}$ is taken into account. The second term in $\mathcal{E}_\text{an}$ with $K^\text{nd}$ being the coefficient of non-diagonal components of the effective anisotropy matrix, reflects the numbering of moments (see Appendix~\ref{sec:discrete}) and the variation of the anisotropy axis within each AFM dimer. The last term in energy~\eqref{eq:en-general}, $\mathcal{E}_\text{f} = - (2/\varsigma)\vec{m}\cdot\vec{h}$ represents the interaction of magnetic moments with the normalized external magnetic field $\vec{h} = \vec{H}/H_0$, where the characteristic field $H_0 = (S/\mu_\textsc{b})\sqrt{|\mathscr{J}\mathscr{K}|}$ and  $\mu_\textsc{b}$ is the Bohr magneton. 

The field-driven reorientation phase transitions are primarily determined by the anisotropic properties of the sample. For example, at the micromagnetic level, the dipolar interaction leads to the same hard-tangential anisotropy as the one introduced above. The strength of this anisotropy is independent of the geometry. For convenience, we normalize each energy term in~\eqref{eq:en-general} to this anisotropy, see purple line in Fig.~\ref{fig:phase_d_hz}(e) ($K^\text{ani} = 1$). The exchange energy in curvilinear AFM spin chains provides the chiral DMI-like response whose strength is characterized by $\varkappa$ (orange line in Fig.~\ref{fig:phase_d_hz}(e)), and anisotropic response, which scales as $\varkappa^2$ and induces the easy axis along $\vec{e}_\textsc{b}$, see red line ($K^\text{x}$) in Fig.~\ref{fig:phase_d_hz}(e). The geometry-governed anisotropic contribution with the coefficient $K^\text{nd}$ provides a tilt of the in-plane anisotropy axes, see green line in Fig.~\ref{fig:phase_d_hz}(e) and schematics in Fig.~\ref{fig:phase_d_hz}(d).

\section{Reorientation phase transitions in magnetic field applied along the ring axis}
\label{sec:perpendicular}

\begin{figure}
\includegraphics[width=\linewidth]{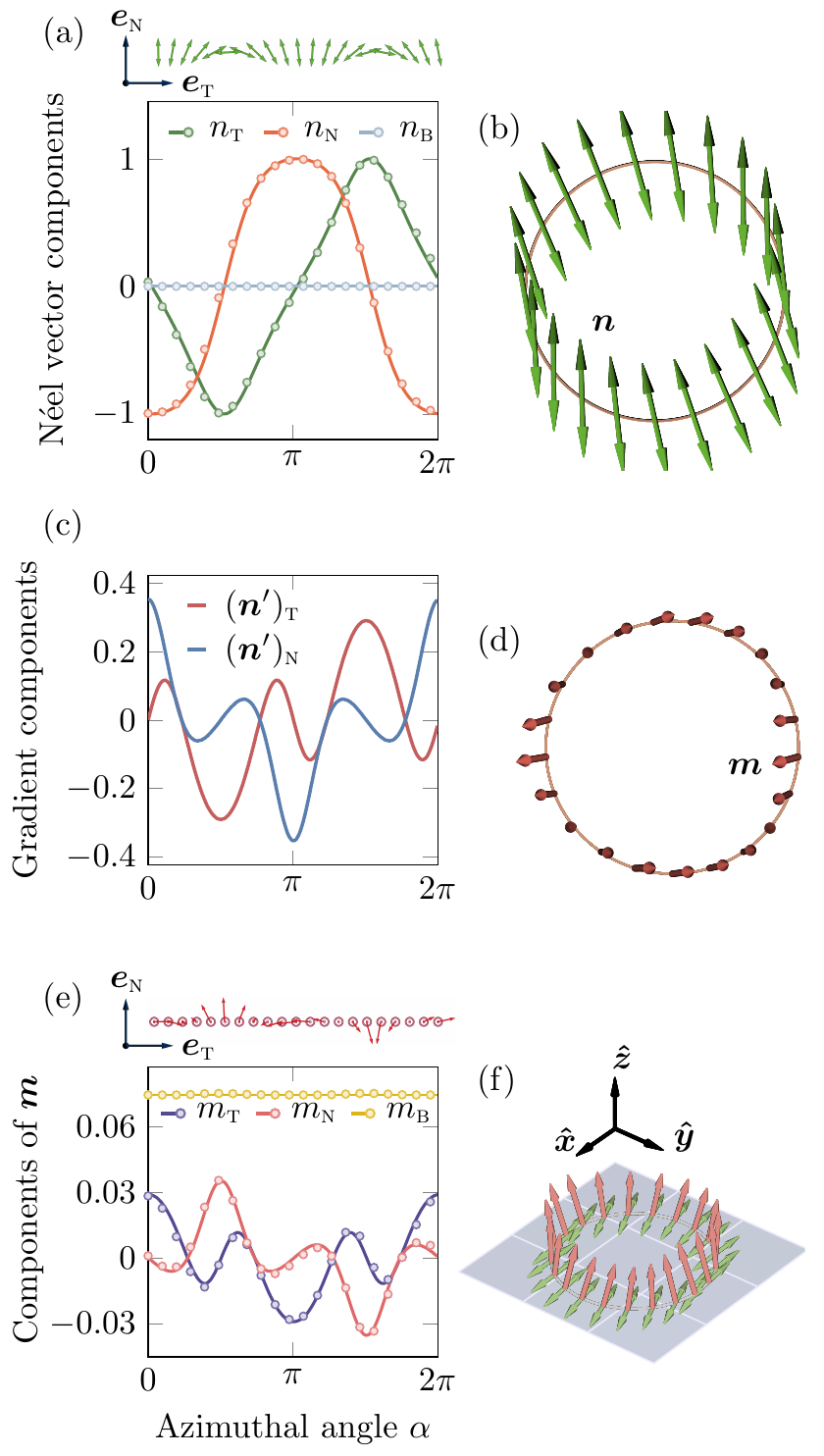}
\caption{\textbf{Onion state in the ring with curvature $\varkappa = 0.75$ exposed to the magnetic field} $\vec{h} = 0.74\vec{e}_\textsc{b}$.
		(a)~Components of the N\'{e}el vector $\vec{n}$.
		(b)~Schematics of the N\'{e}el vector $\vec{n}$ for the onion texture.
		(c)~Gradient components of $\vec{n}$.
		(d)~Schematics of the magnetization vector $\vec{m}$ for the onion texture. 
		(e)~Components of the magnetization vector $\vec{m}$.
		(f)~Superimposed N\'{e}el and magnetization vectors (not to scale).
		In panels (a) and (e) symbols represent data obtained from spin-lattice simulations, solid lines correspond to expressions ~\eqref{eq:onion-n} and~\eqref{eq:onion-m}. 
	}\label{fig:onion_st}
\end{figure} 

\begin{figure}
\includegraphics[width=\linewidth]{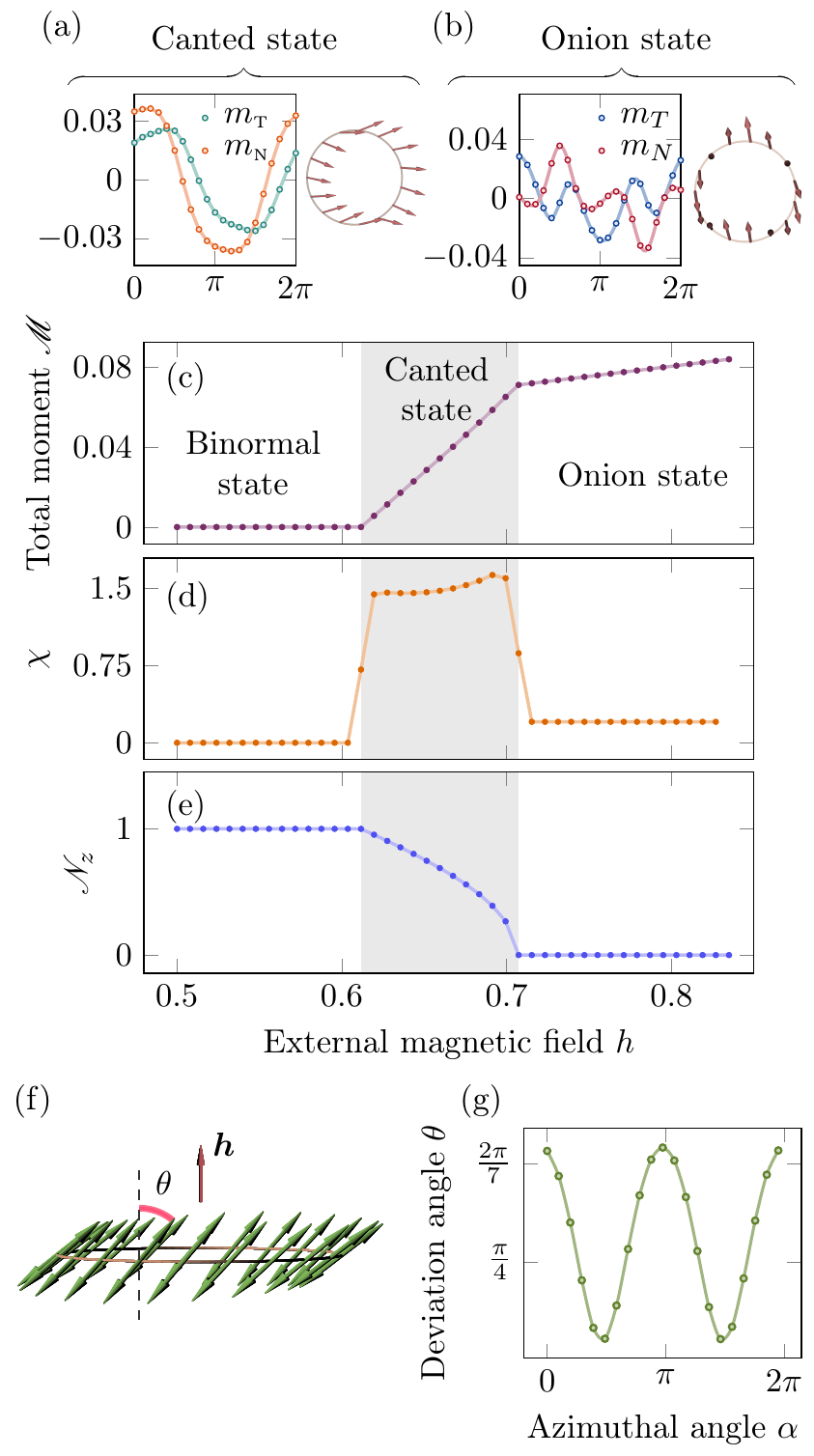}
\caption{\textbf{Canted state in fields} $\vec{h} \parallel \vec{e}_\textsc{b}$. 
		(a)~Components of the magnetization vector $\vec{m}$ in the canted state for the ring with $\varkappa = 0.683$ exposed to the magnetic field $h = 0.66$. 
		(b)~Same for the onion state for the ring with $\varkappa = 0.683$, $h = 0.79$. 
		The magnetic field dependence of (c) the equilibrium magnetic moment normalized by the number of magnetic sites,
		(d)~the differential magnetic susceptibility, and 
		(e)~the total $\vec{n}$ component along the field direction $\mathscr{N}_z$ normalized by the number of magnetic sites. 
		(f)~Schematics of distribution of the N\'{e}el vector in the canted state. Dashed line coincides with the direction $\vec{e}_\textsc{b}$. The angle $\theta$ between $\vec{n}(\vec{r})$ and $\vec{h}$ is indicated as well. Its spatial distribution is shown in panel (g). In all panels, symbols correspond to the results of simulations, lines are guides to the eye.
	}
	\label{fig:canted-state}
\end{figure} 

Fig.~\ref{fig:phase_d_hz}(f) shows spin reorientation phase transitions in AFM rings of different curvature $\varkappa$ exposed to an external magnetic field $\vec{h}$ applied along $\vec{e}_\textsc{b}$ (perpendicular to the ring plane). The diagram includes five different phases. The easy axis of anisotropy stemming from exchange enables the spin-flop transition at field $h_\text{sf}$. Below $h_\text{sf}$, there is the only ground state with $\vec{n}$ oriented along $\vec{e}_\textsc{b}$ (binormal state, blue region in Fig.~\ref{fig:phase_d_hz}(f))~\cite{Pylypovskyi20}. Fields stronger than $h_\text{sf}$ induce the reorientation of $\vec{n}$ and develop a static finite magnetic moment of the ring
\begin{equation}\label{eq:m-via-n}
	\vec{m} = \varsigma\vec{n}\times\left[(\vec{h}-\vec{n}')\times\vec{n}\right]+\mathscr{O}(\varsigma^2)
\end{equation} 
for $h \gtrsim h_\text{sf}$, where the term $\vec{n}'$ is responsible for the geometry-governed weak ferromagnetism~\cite{Pylypovskyi21f}. We find that the field configuration $\vec{h}\parallel \vec{e}_\textsc{b}$ supports two equilibrium states in the spin-flop phase depending on the ring radius. If the sample is sufficiently large (i.e., curvature is less than the critical one discussed below, $\varkappa < \varkappa_c$), the vortex state appears for $h > h^{\text{v}}_\text{sf} = \varkappa$ (pink shaded region in Fig.~\ref{fig:phase_d_hz}(f)), see details in Ref.~\onlinecite{Pylypovskyi21f}. In rings with curvature larger than $\varkappa_c$, there appears a magnetic texture of other symmetry, which is referred to as the onion state (green shaded region in Fig.~\ref{fig:phase_d_hz}(f)). In each ground state in the spin-flop phase, the distribution of the N\'{e}el vector lies in the ring plane. In strong enough fields, the sample becomes completely saturated experiencing the spin-flip transition (brown shaded region in Fig.~\ref{fig:phase_d_hz}(f)). 

In the vicinity of the spin-flop phase transition, the magnetization~\eqref{eq:m-via-n} is small. In this case, the magnetic energy of the ring reads
\begin{equation}\label{eq:en-density-sf}
	E = E_0\int\!\! \mathcal{E}\mathrm{d}\alpha,\quad \mathcal{E} = \vec{n}'^2 + n_\textsc{t}^2 + K^\text{nd} n_\textsc{t}n_\textsc{n} - h^2 +\mathscr{O}(\varsigma^2).
\end{equation}

To find the equilibrium distributions of the order parameters, we parameterize the N\'{e}el vector as $\vec{n} = \vec{e}_\textsc{t} \sin{\vartheta} \cos{\varphi} + \vec{e}_\textsc{n} \sin{\vartheta} \sin{\varphi} + \vec{e}_\textsc{b} \cos{\vartheta}$ with $\vartheta$ and $\varphi$ being the polar and azimuthal angles in the local spherical reference frame. In the spin-flop phase, $\vec{n}$ lies in the ring plane corresponding to $\vartheta = \pi/2$. The spatial distribution of $\vec{n}$ is described by $\varphi(\alpha)$. As a consequence of the curvature-induced tilt of the anisotropy axes from the TNB directions, it is convenient to describe the $\vec{n}(\alpha)$ distribution by measuring the dependence of the angle between $\vec{n}$ and the in-plane anisotropy axis, $\varPhi(\alpha)=\varphi(\alpha) - (1/2) \arctan{2\varsigma\varkappa}$. Minimization of energy~\eqref{eq:en-density-sf} with respect to $\varPhi$ gives the following equation:
\begin{equation}\label{eq:pend-eq}
\varkappa^2 \partial_{\alpha \alpha} \varPhi+\sin{\varPhi}\cos{\varPhi} = 0.
\end{equation}
The vortex state solution with spatially homogeneous distributions of $\vec{m}$ and $\vec{n}$ in the local reference frame is stable for the case of small curvatures. This corresponds to $\varPhi = \pi/2$~\cite{Pylypovskyi21f} and the energy of the vortex state reads ${E_\text{vor}/E_0 = 2\pi(\varkappa^2 - h^2) +\mathscr{O}(\varsigma^2) }$. A spatially inhomogeneous solution of~\eqref{eq:pend-eq}, $\varPhi=-\am\left(x, k\right),$ with $x=\frac{2\ellipticK (k)}{\pi}\alpha$ is referred to as the onion state, which is characterized by the following distribution
\begin{equation} \label{eq:onion-n}
	\vec{n} = \vec{e}_\textsc{t}\cn(x,k)-\vec{e}_\textsc{n}\sn(x,k) +\mathscr{O}(\varsigma),
\end{equation} 
where $\cn(\bullet,k)$ and $\sn(\bullet,k)$ are the elliptic cosine and sine with modulus $k$, respectively~\cite{NIST10}. In this expression, it is taken into account that opposite directions of $\vec{n}$ are physically equivalent. The ring geometry imposes the boundary condition on $\varPhi$, from which the value of $k$ for the given curvature is determined
\begin{equation} \label{eq:onion-bc}
2 \varkappa k \ellipticK(k)=\pi,
\end{equation} 
where $\ellipticK (k)$ is the complete elliptic integral of the first kind~\cite{NIST10}. This gives the energy of the onion texture to be equal ${E_\text{on}/E_0 = 8\varkappa \ellipticE(k)/k+2 \pi\left(1-1/k^2-\varkappa^2-h^2\right)}+\mathscr{O}(\varsigma)$ with $\ellipticE (k)$ being the complete elliptic integral of the second kind~\cite{NIST10}. The critical curvature $\varkappa_c  \approx 0.657$ separates the vortex and onion states. Its value is the solution of the equation $E_\text{vor}(\varkappa) = E_\text{on}(\varkappa)$. The boundary between these states in simulations is $\varkappa_c^\text{num} = 0.669\pm0.014$ and remains constant up to strong fields of $h\sim7$. We note that for ferromagnetic rings in the absence of the external magnetic field, a similar geometrical phase transition between textures of different symmetry is observed with the critical curvature close to $\varkappa_c$~\cite{Sheka15}.

The presence of large spatial derivatives of the components of the N\'{e}el vector [Fig.~\ref{fig:onion_st}(c)] along the tangential direction in the onion state is reflected in the appearance of the
local magnetization. Substituting the expression~\eqref{eq:onion-n} in~\eqref{eq:m-via-n}, the local direction of the magnetization vector reads
\begin{equation}\label{eq:onion-m}
	\begin{aligned}
		\vec{m} 
		=&\pm\varsigma \left[\frac{1}{k}\dn(x,k) - \varkappa \right] [\vec{e}_\textsc{t}\sn(x,k)
		+\vec{e}_\textsc{n}\cn(x,k)] \\
		&+\vec{e}_\textsc{b} \varsigma h +\mathscr{O}(\varsigma^2),
	\end{aligned}
\end{equation} 
see Fig.~\ref{fig:onion_st}(d) and~\ref{fig:onion_st}(e). In contrast to the spin-flop vortex state
with the locally homogeneous magnetization~\cite{Pylypovskyi21f}, the local magnetic moment of the onion state is non-uniform 
and depends on $\varkappa$, c.f. Fig.~\ref{fig:onion_st}(c) and~\ref{fig:onion_st}(e). The binormal component of $\vec{m}$ is determined by $\vec{h}$. The tangential and normal projections of $\vec{m}$ are even and odd functions of $\alpha$, respectively. We note that the energy is degenerate with respect to the sign change of $\vec{n}$ as well as the in-plane components of $\vec{m}$. Schematics of both order parameters in rings is shown in Fig.~\ref{fig:onion_st}(f).

The transition to the spin-flop phase from the uniform ground state in bulk AFMs may occur either as the first-order phase transition with a jump-like change of magnetization along the external field direction, or as two second-order transitions through the so-called canted phase. For a bulk AFM with a spatially homogeneous texture below and above the spin-flop transition, an appearance of the canted phase can be a consequence of the interplay between the exchange and single-ion anisotropy~\cite{Ivanov05a}, higher-order anisotropy terms (e.g., cubic in addition to the uniaxial anisotropy)~\cite{Bogdanov07} or DMI~\cite{Medvedovskaya85, Thio90,Tsukada01}. In the case of ring-shaped AFM spin chains, the formation of the onion state is assured by the curvature-induced DMI with the energy density~$\mathcal{E}_\textsc{dm}=\vec{d}\cdot[\vec{n}'\times\vec{n}]$ and the Dzyaloshinskii vector $\vec{d} = 2\varkappa \vec{e}_\textsc{b}$. Being the part of expression for $\mathcal{E}_\text{x}$, the curvature-induced DMI allows rotation of $\vec{n}$ in the ring plane, see orange line in Fig.~\ref{fig:phase_d_hz}(e)~\cite{Pylypovskyi20}. The canted state in AFM rings is observed for curvatures $\varkappa > \varkappa_\text{cant} \approx 0.585$, see yellow shaded region in Fig.~\ref{fig:phase_d_hz}(f) and details of the state compared with the onion in Fig.~\ref{fig:canted-state}(a) and~\ref{fig:canted-state}(b). We associate the appearance of the canted state with the competition between the Zeeman energy term $\mathcal{E}_\text{f}$, curvature-induced easy-axis anisotropy $K^\text{x}$ and DMI $\mathcal{E}_\textsc{dm}$. In this phase, the total magnetic moment of the ring in the field direction normalized by the number of magnetic sites, $\mathscr{M}$,
grows with the external field faster than in the spin-flop phase, see Fig.~\ref{fig:canted-state}(c) for magnetization and Fig.~\ref{fig:canted-state}(d) for the differential susceptibility $\chi = \partial_h \mathscr{M}$. This is accompanied by a non-linear decay of the total N\'{e}el vector $\mathscr{N}_z$ measured along $\vec{h}$, see Fig.~\ref{fig:canted-state}(e). The canted state is spatially inhomogeneous due to the competition between the spatially varying anisotropy axis $\vec{e}_\text{ani}$ and homogeneous magnetic field $\vec{h}$, see schematics in Fig.~\ref{fig:canted-state}(f) and the distribution of the angle between $\vec{n}$ and $\vec{h}$, $\theta$,  along the azimuthal angle around the ring in Fig.~\ref{fig:canted-state}(g).

Being exposed to the magnetic field of the order of the exchange field, $h \sim h_\text{x} = 1/\varsigma$, spin chains experience the spin-flip phase transition after which the magnetic ordering becomes ferromagnetic (brown shaded region in Fig.~\ref{fig:phase_d_hz}(f)). The N\'{e}el vector vanishes and the total magnetic moment of the ring in the field direction reaches its saturation. Within the limit of the analytical model, which is linear with respect to $\varsigma$, the spin-flip field $h_{\text{spin-flip}} = h_\text{x}$. The critical transition fields, obtained from spin-lattice simulations (Appendix~\ref{sec:simulations}), are reduced up to $0.4\%$ of $h_\text{x}$, see the $h_{\text{spin-flip}}$ line in Fig~\ref{fig:phase_d_hz}(f). 
We associate this small deviation with effects in $\mathscr{O}(\varsigma)$. 
The in-plane magnetization components, which emerged from $\vec{n}'$, produce local spatial modulation of the length of the magnetization vector in the onion state up to $12\%$ in the vicinity of the spin-flop transition. The amplitude of this modulation decays with the increase of the field and is beyond the linear theory for the high-field states.

Approaching the spin-flip field, the critical curvature separating the spin-flop vortex and onion phases is reduced to $\varkappa_c^\text{flip} \approx 0.582$, see Fig~\ref{fig:phase_d_hz}(f). We attribute this change to the curvature-induced DMI similarly to the appearance of the canted state near the vortex phase.

\section{Reorientation phase transitions in tilted fields}
\label{sec:tilted}

\begin{figure*}
\includegraphics[width=\linewidth]{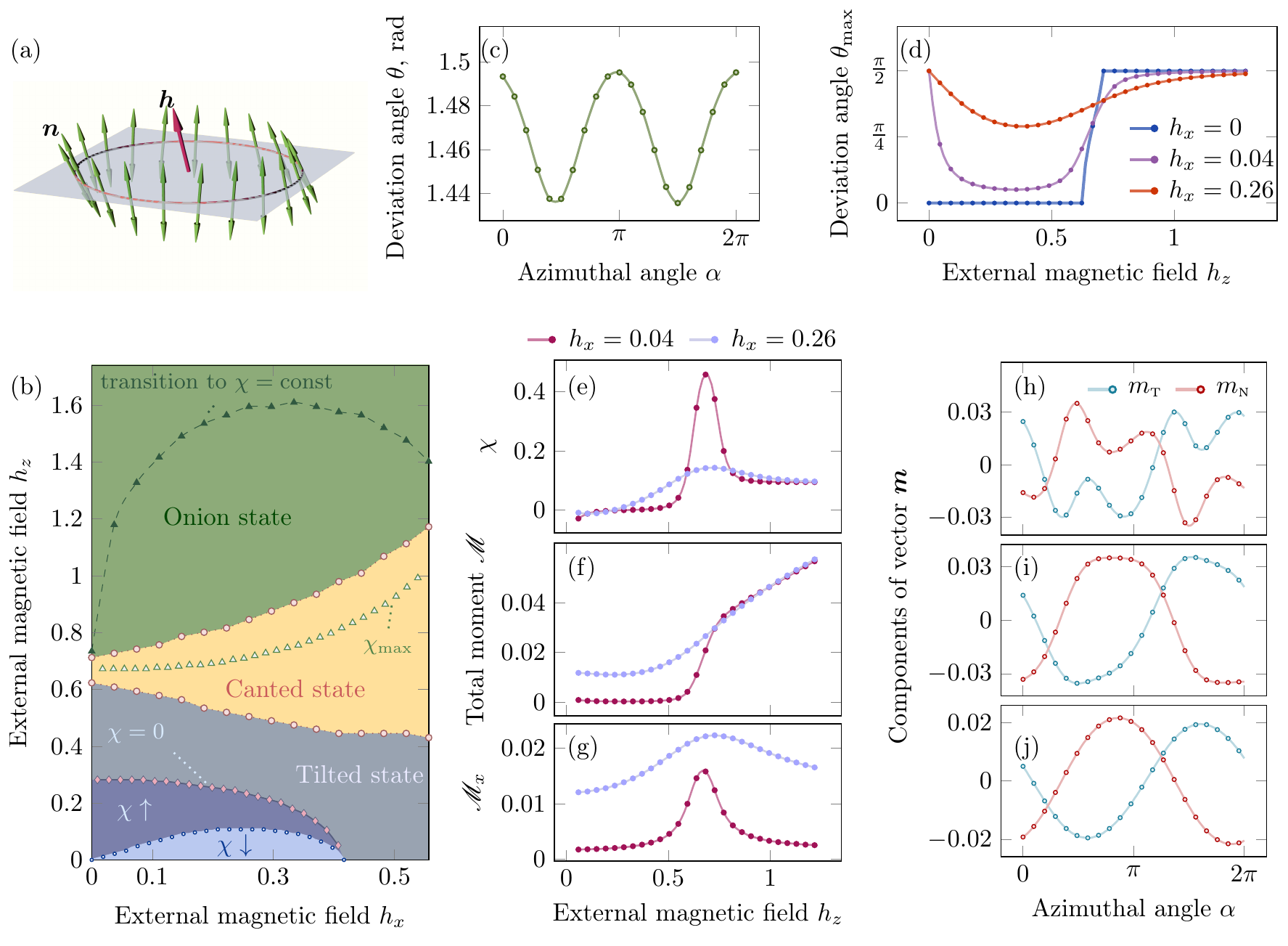}
\caption{\textbf{AFM ring of curvature $\varkappa=0.75 >\varkappa_c$ exposed to the magnetic field $\vec{h}$ tilted from the easy axis.} 
    (a) Schematics of the distribution of the N\'{e}el vector in the tilted state. 
    (b) Spatial distribution of $\theta$ for $h_x=0.04$ and $h_z = 0.83$. 
    (c) Maximum angle $\theta_\text{max}$ between the N\'{e}el vector and field direction as function of $h_z$ for fixed~$h_x$.
    (d) The diagram of equilibrium states for different field directions. Simulations are performed with steps $\Delta h_x = 0.037$ and $\Delta h_z = 0.015$. 
    (e) Differential susceptibility $\chi$ as function of the field component $h_z$ shown for two different values of $h_x$. 
    The magnetic field dependence of (f) the total magnetic moment normalized by the number of sites and (g) the magnetic moment along $\vec{\hat{x}}$ axis for the same values of $h_x$ as used in panel~(e).
    The change of the components of the magnetization vectors with the azimuthal angle $\alpha$ for the (h) onion, (i) canted, and (j) tilted states.
    In all panels, lines are guides to the eye and symbols correspond to simulation results. In panels (e--g), every third symbol from simulations is shown. 
}
\label{fig:ph_onionti}
\end{figure*}

A finite external field applied under a certain angle to the ring axis, $\vec{h}=\vec{\hat{x}} h_x+\vec{\hat{z}} h_z$, breaks the rotational symmetry of the system, see Fig.~\ref{fig:ph_onionti}(a). In uniaxial bulk AFMs (chiral and achiral), this can lead to the appearance of metastable states confined within the astroid-shaped region on the diagram of states in $(h_x,h_z)$ coordinates~\cite{Ivanov05a,Bogdanov07,Medvedovskaya85}. In a curvilinear AFM, lowering of the system symmetry is accompanied by the lifting up of the translational symmetry of the system due to the spatial inhomogeneity of $\vec{e}_\text{ani}$. In the following, we discuss the field-induced states of a ring with curvature $\varkappa > \varkappa_c$ being in the onion state if $\vec{h} \parallel \vec{e}_\textsc{b}$. The field-induced transformation of the spin-flop vortex state with the field $\vec{h}=\vec{\hat{x}} h_x+\vec{\hat{z}} h_z$ is presented in Appendix~\ref{sec:vortex}.

The ($h_x$, $h_z$) diagram of equilibrium states in the tilted magnetic field for the ring of curvature $\varkappa = 0.75$ is shown in Fig.~\ref{fig:ph_onionti}(b). The equilibrium states and their stability regions are determined by analyzing the dependence of the differential susceptibility $\chi$ on $h_z$ measured along the field direction for different $h_x$ [Fig.~\ref{fig:ph_onionti}(e)].

Even below the spin-flop transition, the tilted field develops a component of the total magnetic moment along $\vec{h}$ supplemented by the spatially inhomogeneous texture for the N\'{e}el vector. At low $h_z$, the state which is referred to as the tilted one is developed from the binormal state and is similar to the state in the canted phase (c.f.~Figs.~\ref{fig:canted-state}(f) and~\ref{fig:ph_onionti}(a)). Reflecting the symmetry of a ring, a representative distribution of the deviation angle of $\vec{n}$ with $\alpha$ has two pronounced minima, see Fig.~\ref{fig:ph_onionti}(c). The maximum deviation from the field axis $\theta_\text{max}$ is plotted as function of $h_z$ for different $h_x$ in Fig.~\ref{fig:ph_onionti}(d). It depends on the relation between $h_x$ and $h_z$. If $h_z \lesssim h_x$ keeping the ring in the tilted state, the deviation angle is close to $\pi/2$. The $\theta_\text{max}$ decreases with the increase of $h_z$ and reaches a fixed value for the given $h_x$. In the tilted state, the total moment $\mathscr{M}$ initially decreases because of the effective anisotropy along $\vec{e}_\textsc{b}$ [Fig.~\ref{fig:ph_onionti}(f)]. The respective regions of negative susceptibility are shown in Fig.~\ref{fig:ph_onionti}(b) with light blue (decreasing negative $\chi$ with increasing $h_z$) and dark blue (increasing non-positive $\chi$ with increasing $h_z$). The locus of fields corresponding to $\chi = 0$ is shown with open rhombuses. 

The region of the canted phase expands into regions of the tilted and onion states with larger $h_x$, see Fig.~\ref{fig:ph_onionti}(b). The boundaries of the canted state in Fig.~\ref{fig:ph_onionti}(b) are determined by the inflection points for the susceptibility $\chi(h_z)$. Proceeding to the canted state by the increase of the $h_z$ is characterized by the increase of $\theta_{\text{max}}$. The amplitude of the spatial distribution of the angle between $\vec{n}$ and $\vec{h}$ ($\theta_{\text{max}} - \theta_{\text{min}}$)
reaches its maximum in the canted state (up to $\approx 0.2$ rad for small $h_x$)
and becomes smaller with $\theta_{\text{max}}$ approaching $\pi/2$ in the onion state. Open triangles in Fig.~\ref{fig:ph_onionti}(b) correspond to values of $h_x$ and $h_z$ at which the differential susceptibility $\chi$ reaches its maximum [Fig.~\ref{fig:ph_onionti}(b) and~\ref{fig:ph_onionti}(g)]. The tilted and canted states are similar with respect to the distribution of the magnetization vector, c.f. Fig.~\ref{fig:ph_onionti}(i) and~\ref{fig:ph_onionti}(j). Still, we note that the mechanism of their stabilization reflected in the behavior of the magnetic susceptibility is different. The onion state in the tilted field is qualitatively similar to the one discussed for the field along $\vec{e}_\textsc{b}$, c.f. Fig.~\ref{fig:ph_onionti}(h) and Fig.~\ref{fig:onion_st}(e).

The change of the total magnetic moment of the ring in a tilted field is shown in Fig.~\ref{fig:ph_onionti}(f) and~\ref{fig:ph_onionti}(g). There are three well-distinguished phases for sufficiently small $h_x$, namely almost constant $\mathscr{M}$, rapidly growing $\mathscr{M}$ and slowly growing $\mathscr{M}$. The transition between phases becomes smoother with an increase of $h_x$. The component of magnetic moment along the $\vec{\hat{x}}$ direction has a pronounced maximum within the canted state, see Fig.~\ref{fig:ph_onionti}(g).

\section{Discussion}
\label{sec:summary}

To summarize, we describe field-induced reorientation transitions of the N\'{e}el vector $\vec{n}$ in an AFM spin chain consisting of even number of spins arranged in a ring. The model accounts for the isotropic exchange, hard-axis anisotropy with the axis tangential to the ring and Zeeman interaction. Having intrinsic hard-axis anisotropy, such rings have a ground state with the N\'{e}el vector perpendicular to the ring's plane independent of curvature in absence of external magnetic fields~\cite{Pylypovskyi20}. The critical fields, characteristic of the spin reorientation transitions, are determined by the curvature. The spin-flop state consists of the vortex and onion phases for rings of small and large curvatures, respectively. The spin-flop transition behaves as the first-order one for rings with curvatures $\varkappa < \varkappa_\text{cant} \approx 0.585$. Otherwise, the transition is of the second order and happens via the intermediate canted state. The region of the canted state is expanded in fields applied under an angle to the ring axis. Approaching the spin-flip transition, the critical curvature between the vortex and onion states decreases to $\varkappa_c^\text{flip} \approx 0.582$. The onion state is characterized by a weak ferromagnetic response. The strength of the weak ferromagnetism is determined by $\vec{n}'$. The respective moment lies in the plane of $\vec{n}'$.

This work provides the first insight into the influence of the geometric curvature on the spin reorientation transitions induced by external magnetic fields in curvilinear AFM spin chains. A spin chain arranged along plane curves is a paradigmatic example, which allows to follow the phase transitions driven by the geometry for the case when the intrinsic anisotropy has its hard axis along the tangential direction. This hard axis anisotropy can be modified by or originate from the dipolar interaction~\cite{Pylypovskyi20}. Our results can be used to analyze the spin-flop transitions for other geometries and types of anisotropy, where the ground state is not necessarily uniform. 
We note that monitoring the curvature-dependent critical field, which is needed to induce the change of the magnetic state, provides a complementary method to determine material parameters for low-dimensional AFMs based on molecular magnets~\cite{Sorolla19, Fu21}, DNA-based systems~\cite{Mizoguchi07,Kesama18} or fabricated by means of atom-by-atom engineering~\cite{Loth12, Khajetoorians12}. 

The strength of the AFM exchange coupling in spin chains varies over several orders of magnitude, e.g. $\sim 10^{-24}$\,J
for Cu chains with monochloride bridges~\cite{Santana12}, 
$\sim 10^{-23}$\,J for pyridine-based Cu chains~\cite{Drahos15} and $\sim 10^{-22}$\,J for molecular wheels \ch{Cr8Cd}~\cite{Furukawa08,Guidi15,Garlatti20}. The latter is about an order of magnitude smaller than the exchange in the bulk 
\ch{Cr2O3}
($\sim 10^{-21}$\,J)~\cite{Kota16}. Assuming that a certain nanomagnet possesses the hard-axis intrinsic anisotropy with the strength of the same order as reported for \ch{Cr8Cd}~\cite{Guidi15,Furukawa08}, it is possible to estimate the magnetic length $\ell$ to be about 5 to 10 lattice constants and characteristic fields $H_0$ to be about 3.5\,T, which is readily achievable in laboratory experiments.
Such a molecular ring with eight atoms is expected to be in the spin-flop onion state. Parameters of the spin-flop transition can be also tuned in systems prepared by atom-by-atom engineered relying on the proper selection of the substrate and spin-carrying atoms~\cite{Hirjibehedin07}. Further experimental investigations of these systems and the comparison with the theoretical predictions of this work should also give insight into the role of quantum effects in curvilinear low-dimensional AFMs. Lowering of the spin-flop and spin-flip fields in spin chains due to strong anisotropy, in comparison with bulk AFMs, paves the way towards reconfigurable spintronic devices, whose operational modes are different below and above the spin-flop transition.

The influence of the curvature-induced
DMI on the spin-flop transition requires further analysis. In particular, it is insightful to compare the results presented above with the antiferromagnets in rolled-up geometries~\cite{Ueltzhoeffer16}, which represent a development of the ring geometry into a tube. 
In planar chains, the presence of the Lifshitz invariant proportional to curvature is reflected in the appearance of the canted phase in external field for large ring curvatures. In contrast, AFM spin chains arranged along space curves have two Lifshitz invariants and possess the geometrically-governed helimagnetic phase transition driven by the Lifshitz invariant proportional to torsion in the absence of the field~\cite{Pylypovskyi20}.
The presence of DMI terms of different symmetries in 3D curved spin chains should make the diagram of field-induced states richer. We anticipate that this behavior could be comparable with the spin-flop of a planar ring in the tilted field, since the homogeneous magnetic field does not coincide with the easy axis of a 3D spin chain. 

\begin{acknowledgements}
The authors thank Dr. Nina Elkina (HZDR) for support with simulations. Numerical calculations are performed using the OpenStack and Hemera facilities at the HZDR~\cite{hzdr}. Y.A.B. acknowledges financial support from the UKRATOP-project (funded by the BMBF; reference 01DK18002). This work is financed in part via the German Research Foundation (DFG) under the grants MC 9/22–1, MA 5144/22-1, MA 5144/24-1. 
\end{acknowledgements}

\appendix

\section{Spin-lattice simulations}
\label{sec:simulations}

To perform numerical analysis of curvilinear AFMs, we use the in-house developed SLaSi package~\cite{Pylypovskyi20,SLaSi}. The Landau--Lifshitz--Gilbert equation
\begin{equation}
    \dfrac{\mathrm{d}\vec{\mu}_i}{\mathrm{d}t} = \dfrac{1}{\hbar S} \vec{\mu}_i \times \dfrac{\partial \mathcal{H}}{\partial \vec{\mu}_i} + \alpha_\textsc{g} \dfrac{\mathrm{d}\vec{\mu}_i}{\mathrm{d}t},\quad i = \overline{1,N}
\end{equation}
is solved numerically to obtain equilibrium magnetization states. Here, $\vec{\mu}_i$ is the unit vector of magnetic moment for $i$-th chain site, $\hbar$ is the reduced Planck constant, $\alpha_\textsc{g}$ is the Gilbert damping, $N$ is the number of sites in the chain. The spin-lattice Hamiltonian reads
\begin{equation}\label{eq:hamiltonian}
    \begin{aligned}
        \mathcal{H} & = \dfrac{\mathscr{J}S^2}{2} \left( \sum_{i=1}^{N-1} \vec{\mu}_i\cdot \vec{\mu}_{i+1} + \vec{\mu}_{N}\cdot \vec{\mu}_1\right) \\
        & +  \dfrac{\mathscr{K}S^2}{2} \sum_{i=1}^{N} (\vec{\mu}_i\cdot \vec{e}_\textsc{t}^i)^2 - 2\mu_\textsc{b}S\sum_{i=1}^{N} \vec{\mu}_i\cdot \vec{H}.
    \end{aligned}
\end{equation}
The length of all magnetic moments is the same, $|\vec{\mu}_i| = 1$, $i = \overline{1,N}$ and $\vec{e}_\textsc{t}^i$ is the unit vector determining the tangential direction for $i$-th chain site. The sign of the $\mathscr{J}$ depends on the definition of the exchange part of the Hamiltonian. Here, $\mathscr{J}>0$ favours the antiparallel orientation of $\vec{\mu}_i$, $\vec{\mu}_{i+1}$ in~\eqref{eq:hamiltonian}. For all simulations we used $\alpha_\textsc{g} = 0.5$, $S = 1 $, $a_0 = 0.3$ nm, $\mathscr{J}=\num{1e-22}$ J, $\mathscr{K}=\num{4e-20}$ J
which gives $\ell=5 a_0$ and $\varsigma=0.1$. The system is considered to be in equilibrium if $\max |\mathrm{d}\vec{\mu}_i/\mathrm{d}t| < 10^{-14}$ Hz. 

To determine the preferable equilibrium state for the given magnetic field, the relaxation is done for different initial states: with the N\'{e}el vector distribution being homogeneous in the local or laboratory reference frames, namely states $\vec{n}\parallel\vec{e}_\textsc{b}$, $\vec{n}\parallel\vec{e}_\textsc{n}$ and $\vec{n}\parallel \vec{\hat{x}}$. After relaxation, their resulting distributions and energies are compared. The equilibrium state is considered to be the one possessing the lowest energy. To determine the boundary between the vortex and onion states, as well as boundaries between the states when the ring is exposed to a tilted magnetic field, additional simulations were done with the initial distribution set as in \eqref{eq:onion-n} for the given curvature. 
In the canted state region, every initial distribution relaxes to the canted state (vortex and onion states are not stable in this region). This is in contrast to the case of fields above the canted state, where the onion and vortex states can be metastable ones. The boundary of the region corresponding to the canted state in Fig.~\ref{fig:phase_d_hz}(f) is built based on the results of simulations with $\ell=10 a_0$ for $\mathscr{J}=\num {4e-22}$\,J to provide a denser set of numerically obtained points.

\section{Discrete order parameters}
\label{sec:discrete}

\begin{figure}
\includegraphics[width=\linewidth]{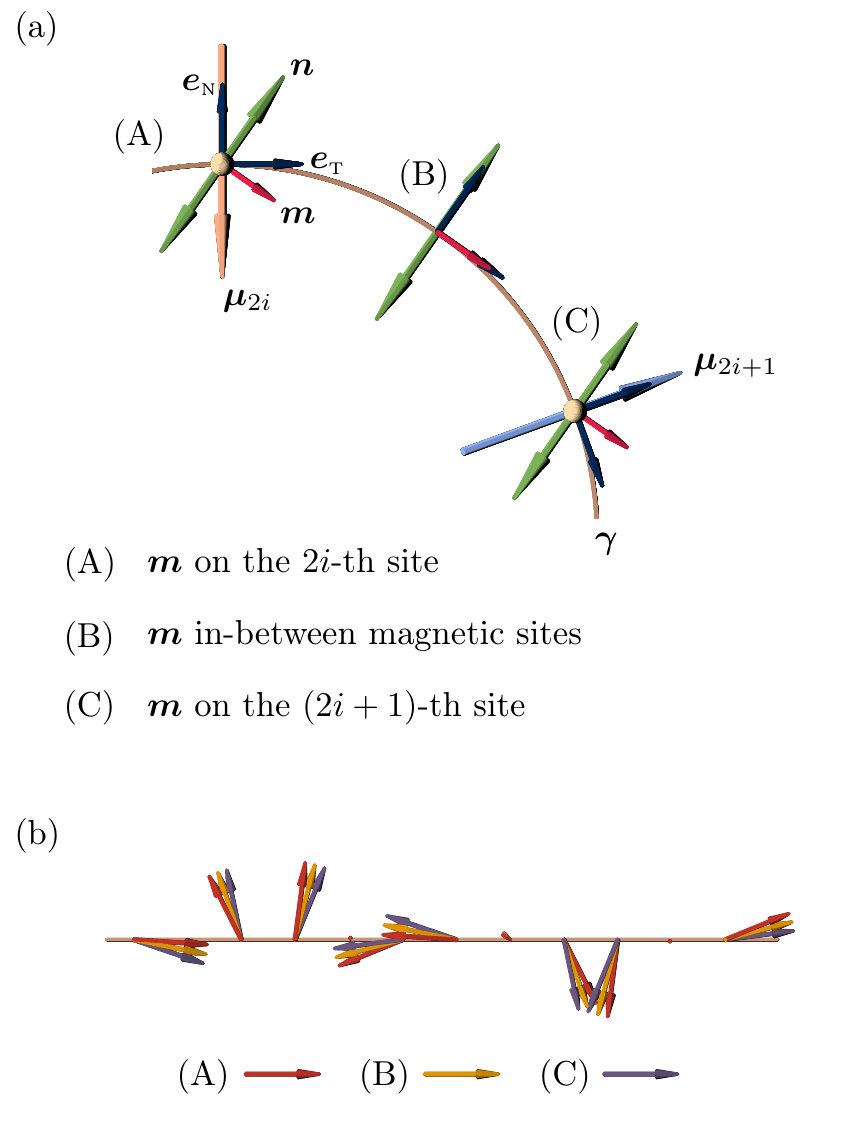}
\caption{\textbf{Dependence of the weak ferromagnetic response on the spatial localization of the discrete order parameter along the curve $\vec{\gamma}$.}
    (a) $i$-th AFM dimer consisting of $\vec{\mu}_{2i}$ and $\vec{\mu}_{2i+1}$ magnetic moments. Three possible orientations of $\vec{n}_i$ and $\vec{m}_i$ are shown for pairs of $\{\vec{n}_i, \vec{m}_i\}$ with centering of the order parameters at (A) $2i$-th site, (B) central point on the curve $\vec{\gamma}$ between the chain sites, and (C) $(2i+1)$-th site.
    (b) The change of the direction of the magnetization vectors for the ring with $\varkappa =1.3$ exposed to the magnetic field $\vec{h} = 0.83 \vec{e}_\textsc{b}$ in the onion state. In both panels, the lengths of the vectors are not to scale. 
    }
\label{fig:discrete}
\end{figure} 

In the continuum description, the N\'{e}el vector $\vec{n}(\vec{r})$ and magnetization vector $\vec{m}(\vec{r})$ are defined at each point $\vec{r}$ of the curve. To compare respective micromagnetic models with spin lattice simulations, it is crucial to take into account that the $i$-th AFM unit cell for spin chains is a dimer $\{\vec{\mu}_{2i}, \vec{\mu}_{2i+1}\}$. Furthermore, the anisotropy axis changes its direction for each of the magnetic moments within this dimer. The anisotropy direction per each magnetic moment is determined by the chain shape via its local atomistic surrounding and coordinates of the neighboring moments. Following the micromagnetic transition from the discrete to continuum models described in Ref.~\onlinecite{Pylypovskyi21f}, in this work, we associate the discrete order parameters $\vec{n}_i = (\vec{\mu}_{2i} - \vec{\mu}_{2i+1})/(2\mu)$ and $\vec{m}_i = (\vec{\mu}_{2i} + \vec{\mu}_{2i+1})/(2\mu)$ with the spatial localization of the moment $\vec{\mu}_{2i}$. This is reflected in $\mathcal{E}_\text{an}$. One can rewrite the energy density of the intrinsic anisotropy as $\vec{n} \hat{K} \vec{n}$, where $\hat{K} = \begin{Vmatrix} 1 & K^{\text{nd}}/2 & 0\\ K^{\text{nd}}/2 & 0 & 0\\ 0 & 0 &0\end{Vmatrix}$~\cite{Pylypovskyi21f}. For the chosen location of the order parameters in the dimer, $K^{\text{nd}}=2 \varsigma \varkappa$. Other possibilities for a dimer placed on the curve $\vec{\gamma}$ are shown in Fig.~\ref{fig:discrete}. For a ring possessing a constant curvature, the point in the geometrical center of the dimer along $\vec{\gamma}$ has the tangential direction $\vec{e}_\textsc{t}^{2i+1/2}$ along the line connecting $2i$-th and $(2i+1)$-th sites ($K^\text{nd} = 0$). This is a special case of high symmetry specific for the ring geometry, which is absent for curves of arbitrary geometry. 

\section{Spin flop vortex phase}
\label{sec:vortex}
\begin{figure*}
\includegraphics[width=\linewidth]{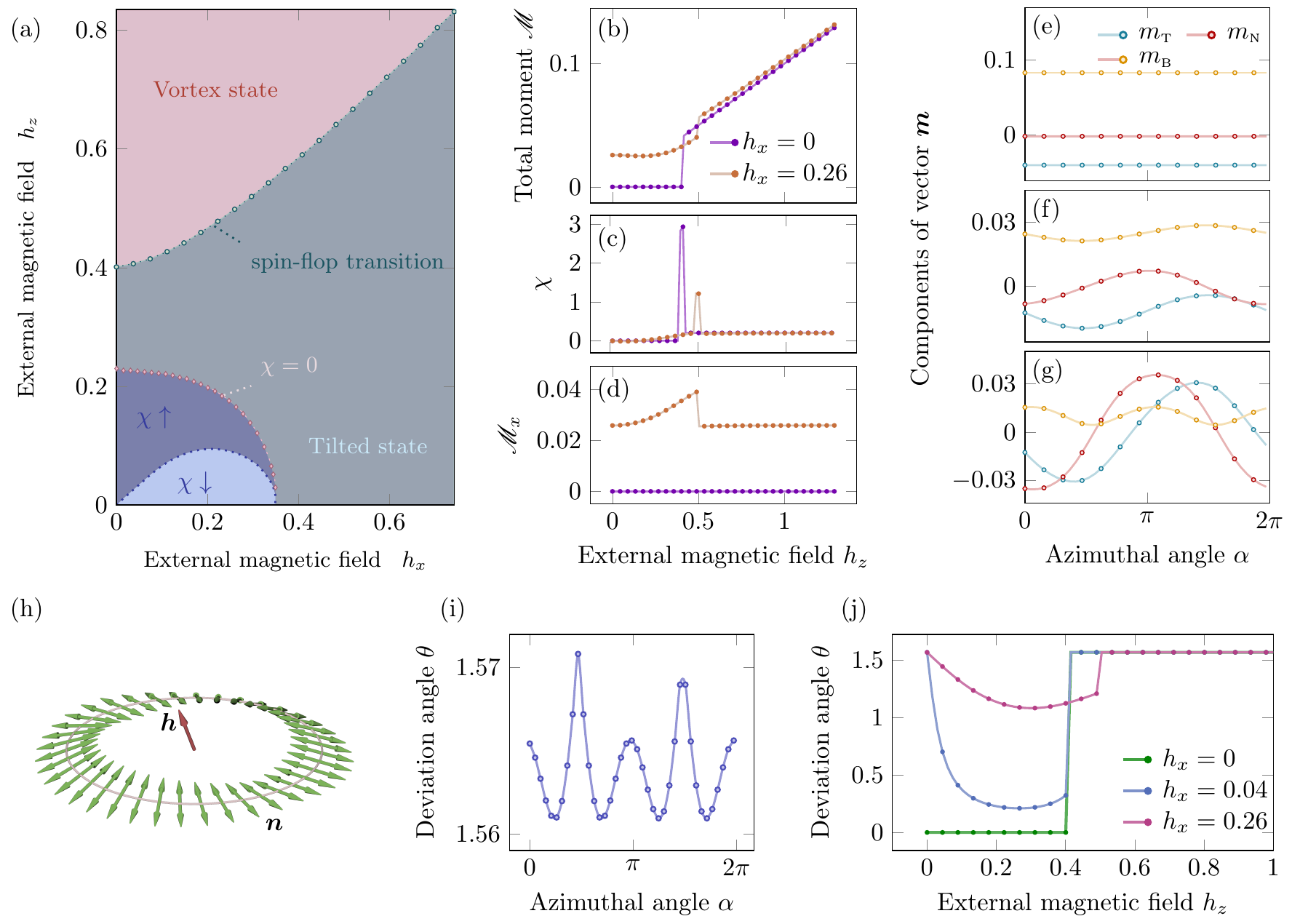}
\caption{\textbf{Spin-flop vortex phase in the ring of curvature} $\varkappa = 0.4<\varkappa_\textrm{cant}$ \textbf{exposed to a tilted external magnetic field.} 
    (a)~The diagram of equilibrium states for different field directions.
    (b)~The total magnetic moment normalized to the number of sites as function of the field $h_z$ for two different values of $h_x$.
    The magnetic field dependence of (c) the differential susceptibility and (d) the magnetic moment along the $\vec{\hat{x}}$ axis for values of $h_x$ as in panel~(b).
    Components of the magnetization vectors for the (e)~vortex state in $\vec{h}\parallel\vec{e}_\textsc{b}$, (f) vortex state in the tilted field, and (g) tilted states.
    (h)~Schematics of the distribution of the N\'{e}el vector in the vortex state for $h_x=0.78$ and $h_z = 1.75$.
    (i)~The spatial distribution of the angle between the N\'{e}el vector and the field direction for $h_x=0.26$ and $h_z = 0.83$.
    (j)~Maximum angle between the N\'{e}el vector and the field direction as function of $h_z$ for fixed $h_x$. 
    In panels (b-g, j), every third symbol from simulations is shown.
}
\label{fig:ph_vortexti}
\end{figure*} 

The diagram of equilibrium states in a tilted field for the ring of curvature $\varkappa < \varkappa_\text{cant}$ possessing the spin-flop vortex state in $h_x=0$ is shown in Fig.~\ref{fig:ph_vortexti}(a). An increase of the total magnetic moment with $h$ reveals the first-order phase transition, see Fig.~\ref{fig:ph_vortexti}(b). The finite jumps in $\chi(h_z)$ in Fig.~\ref{fig:ph_vortexti}(c) are related to the discrete set of points to determine the phase transition by the difference scheme. A jump of the total magnetic moment along $\vec{\hat{x}}$ as function of $h_z$ being the consequence of the first-order transition is clearly seen in Fig.~\ref{fig:ph_vortexti}(d).

For non-zero $h_x$ values, the tilted state is qualitatively similar to the tilted state discussed for $\varkappa>\varkappa_c$. The vortex state in the tilted field [Fig.~\ref{fig:ph_vortexti}(h)] is topologically equivalent to the spin-flop state in $\vec{h}\parallel\vec{e}_\textsc{b}$~\cite{Pylypovskyi21f}, and exists for sufficiently large values of the field tilt angles. The deviation of the N\'{e}el vector from the symmetry plane measured by the angle $\theta$ as in Fig.~\ref{fig:ph_onionti}(a) and~\ref{fig:ph_onionti}(b) is shown in Fig.~\ref{fig:ph_vortexti}(i) and~\ref{fig:ph_vortexti}(j).


\begin{thebibliography}{50}%
\makeatletter
\providecommand \@ifxundefined [1]{%
 \@ifx{#1\undefined}
}%
\providecommand \@ifnum [1]{%
 \ifnum #1\expandafter \@firstoftwo
 \else \expandafter \@secondoftwo
 \fi
}%
\providecommand \@ifx [1]{%
 \ifx #1\expandafter \@firstoftwo
 \else \expandafter \@secondoftwo
 \fi
}%
\providecommand \natexlab [1]{#1}%
\providecommand \enquote  [1]{``#1''}%
\providecommand \bibnamefont  [1]{#1}%
\providecommand \bibfnamefont [1]{#1}%
\providecommand \citenamefont [1]{#1}%
\providecommand \href@noop [0]{\@secondoftwo}%
\providecommand \href [0]{\begingroup \@sanitize@url \@href}%
\providecommand \@href[1]{\@@startlink{#1}\@@href}%
\providecommand \@@href[1]{\endgroup#1\@@endlink}%
\providecommand \@sanitize@url [0]{\catcode `\\12\catcode `\$12\catcode
  `\&12\catcode `\#12\catcode `\^12\catcode `\_12\catcode `\%12\relax}%
\providecommand \@@startlink[1]{}%
\providecommand \@@endlink[0]{}%
\providecommand \url  [0]{\begingroup\@sanitize@url \@url }%
\providecommand \@url [1]{\endgroup\@href {#1}{\urlprefix }}%
\providecommand \urlprefix  [0]{URL }%
\providecommand \Eprint [0]{\href }%
\providecommand \doibase [0]{https://doi.org/}%
\providecommand \selectlanguage [0]{\@gobble}%
\providecommand \bibinfo  [0]{\@secondoftwo}%
\providecommand \bibfield  [0]{\@secondoftwo}%
\providecommand \translation [1]{[#1]}%
\providecommand \BibitemOpen [0]{}%
\providecommand \bibitemStop [0]{}%
\providecommand \bibitemNoStop [0]{.\EOS\space}%
\providecommand \EOS [0]{\spacefactor3000\relax}%
\providecommand \BibitemShut  [1]{\csname bibitem#1\endcsname}%
\let\auto@bib@innerbib\@empty
\bibitem [{\citenamefont {Turov}(1965)}]{Turov65}%
  \BibitemOpen
  \bibfield  {author} {\bibinfo {author} {\bibfnamefont {E.}~\bibnamefont
  {Turov}},\ }\href
  {https://www.amazon.com/Physical-Properties-Magnetically-Ordered-Crystals/dp/0127049509?SubscriptionId=AKIAIOBINVZYXZQZ2U3A&tag=chimbori05-20&linkCode=xm2&camp=2025&creative=165953&creativeASIN=0127049509}
  {\emph {\bibinfo {title} {Physical Properties of Magnetically Ordered
  Crystals}}}\ (\bibinfo  {publisher} {Academic Press},\ \bibinfo {year}
  {1965})\BibitemShut {NoStop}%
\bibitem [{\citenamefont {Ivanov}(2005)}]{Ivanov05a}%
  \BibitemOpen
  \bibfield  {author} {\bibinfo {author} {\bibfnamefont {B.~A.}\ \bibnamefont
  {Ivanov}},\ }\bibfield  {title} {\bibinfo {title} {Mesoscopic
  antiferromagnets: statics, dynamics, and quantum tunneling ({R}eview)},\
  }\href {https://doi.org/10.1063/1.2008127} {\bibfield  {journal} {\bibinfo
  {journal} {Low Temperature Physics}\ }\textbf {\bibinfo {volume} {31}},\
  \bibinfo {pages} {635} (\bibinfo {year} {2005})}\BibitemShut {NoStop}%
\bibitem [{\citenamefont {Šmejkal}\ \emph {et~al.}(2022)\citenamefont
  {Šmejkal}, \citenamefont {MacDonald}, \citenamefont {Sinova}, \citenamefont
  {Nakatsuji},\ and\ \citenamefont {Jungwirth}}]{Smejkal22}%
  \BibitemOpen
  \bibfield  {author} {\bibinfo {author} {\bibfnamefont {L.}~\bibnamefont
  {Šmejkal}}, \bibinfo {author} {\bibfnamefont {A.~H.}\ \bibnamefont
  {MacDonald}}, \bibinfo {author} {\bibfnamefont {J.}~\bibnamefont {Sinova}},
  \bibinfo {author} {\bibfnamefont {S.}~\bibnamefont {Nakatsuji}},\ and\
  \bibinfo {author} {\bibfnamefont {T.}~\bibnamefont {Jungwirth}},\ }\bibfield
  {title} {\bibinfo {title} {Anomalous hall antiferromagnets},\ }\href
  {https://doi.org/doi.org/10.1038/s41578-022-00430-3} {\bibfield  {journal}
  {\bibinfo  {journal} {Nature Reviews Materials}\ ,\ \bibinfo {pages} {2058}}
  (\bibinfo {year} {2022})}\BibitemShut {NoStop}%
\bibitem [{\citenamefont {Baltz}\ \emph {et~al.}(2018)\citenamefont {Baltz},
  \citenamefont {Manchon}, \citenamefont {Tsoi}, \citenamefont {Moriyama},
  \citenamefont {Ono},\ and\ \citenamefont {Tserkovnyak}}]{Baltz18}%
  \BibitemOpen
  \bibfield  {author} {\bibinfo {author} {\bibfnamefont {V.}~\bibnamefont
  {Baltz}}, \bibinfo {author} {\bibfnamefont {A.}~\bibnamefont {Manchon}},
  \bibinfo {author} {\bibfnamefont {M.}~\bibnamefont {Tsoi}}, \bibinfo {author}
  {\bibfnamefont {T.}~\bibnamefont {Moriyama}}, \bibinfo {author}
  {\bibfnamefont {T.}~\bibnamefont {Ono}},\ and\ \bibinfo {author}
  {\bibfnamefont {Y.}~\bibnamefont {Tserkovnyak}},\ }\bibfield  {title}
  {\bibinfo {title} {Antiferromagnetic spintronics},\ }\href
  {https://doi.org/10.1103/RevModPhys.90.015005} {\bibfield  {journal}
  {\bibinfo  {journal} {Reviews of Modern Physics}\ }\textbf {\bibinfo {volume}
  {90}},\ \bibinfo {pages} {015005} (\bibinfo {year} {2018})}\BibitemShut
  {NoStop}%
\bibitem [{\citenamefont {Gomonay}\ \emph {et~al.}(2017)\citenamefont
  {Gomonay}, \citenamefont {Jungwirth},\ and\ \citenamefont
  {Sinova}}]{Gomonay17}%
  \BibitemOpen
  \bibfield  {author} {\bibinfo {author} {\bibfnamefont {O.}~\bibnamefont
  {Gomonay}}, \bibinfo {author} {\bibfnamefont {T.}~\bibnamefont {Jungwirth}},\
  and\ \bibinfo {author} {\bibfnamefont {J.}~\bibnamefont {Sinova}},\
  }\bibfield  {title} {\bibinfo {title} {Concepts of antiferromagnetic
  spintronics},\ }\href {https://doi.org/10.1002/pssr.201700022} {\bibfield
  {journal} {\bibinfo  {journal} {physica status solidi ({RRL}) - Rapid
  Research Letters}\ }\textbf {\bibinfo {volume} {11}},\ \bibinfo {pages}
  {1700022} (\bibinfo {year} {2017})}\BibitemShut {NoStop}%
\bibitem [{\citenamefont {Turov}\ \emph {et~al.}(2001)\citenamefont {Turov},
  \citenamefont {Kolchanov}, \citenamefont {Menshenin}, \citenamefont
  {Mirsayev},\ and\ \citenamefont {Nikolaev}}]{Turov01en}%
  \BibitemOpen
  \bibfield  {author} {\bibinfo {author} {\bibfnamefont {E.~A.}\ \bibnamefont
  {Turov}}, \bibinfo {author} {\bibfnamefont {A.~V.}\ \bibnamefont
  {Kolchanov}}, \bibinfo {author} {\bibfnamefont {V.~V.}\ \bibnamefont
  {Menshenin}}, \bibinfo {author} {\bibfnamefont {I.~F.}\ \bibnamefont
  {Mirsayev}},\ and\ \bibinfo {author} {\bibfnamefont {V.~V.}\ \bibnamefont
  {Nikolaev}},\ }\href@noop {} {\emph {\bibinfo {title} {Symmetry and physical
  properties of antiferromagnets}}}\ (\bibinfo  {publisher} {FIZMATLIT},\
  \bibinfo {address} {Moscow},\ \bibinfo {year} {2001})\BibitemShut {NoStop}%
\bibitem [{\citenamefont {Kosevich}\ \emph {et~al.}(1990)\citenamefont
  {Kosevich}, \citenamefont {Ivanov},\ and\ \citenamefont
  {Kovalev}}]{Kosevich90}%
  \BibitemOpen
  \bibfield  {author} {\bibinfo {author} {\bibfnamefont {A.~M.}\ \bibnamefont
  {Kosevich}}, \bibinfo {author} {\bibfnamefont {B.~A.}\ \bibnamefont
  {Ivanov}},\ and\ \bibinfo {author} {\bibfnamefont {A.~S.}\ \bibnamefont
  {Kovalev}},\ }\bibfield  {title} {\bibinfo {title} {Magnetic solitons},\
  }\href {https://doi.org/10.1016/0370-1573(90)90130-T} {\bibfield  {journal}
  {\bibinfo  {journal} {Physics Reports}\ }\textbf {\bibinfo {volume} {194}},\
  \bibinfo {pages} {117} (\bibinfo {year} {1990})}\BibitemShut {NoStop}%
\bibitem [{\citenamefont {Borisov}\ and\ \citenamefont
  {Kiselev}(2011)}]{Borisov11a}%
  \BibitemOpen
  \bibfield  {author} {\bibinfo {author} {\bibfnamefont {A.~B.}\ \bibnamefont
  {Borisov}}\ and\ \bibinfo {author} {\bibfnamefont {V.~V.}\ \bibnamefont
  {Kiselev}},\ }\href@noop {} {\emph {\bibinfo {title} {Nonlinear waves,
  solitons and localized structures in magnets. Vol.2}}}\ (\bibinfo
  {publisher} {Russian Academy of Science (in russian)},\ \bibinfo {year}
  {2011})\BibitemShut {NoStop}%
\bibitem [{\citenamefont {Pisarev}\ \emph {et~al.}(1997)\citenamefont
  {Pisarev}, \citenamefont {Fiebig},\ and\ \citenamefont
  {Fr\"{o}hlich}}]{Pisarev97}%
  \BibitemOpen
  \bibfield  {author} {\bibinfo {author} {\bibfnamefont {R.~V.}\ \bibnamefont
  {Pisarev}}, \bibinfo {author} {\bibfnamefont {M.}~\bibnamefont {Fiebig}},\
  and\ \bibinfo {author} {\bibfnamefont {D.}~\bibnamefont {Fr\"{o}hlich}},\
  }\bibfield  {title} {\bibinfo {title} {Nonlinear optical spectroscopy of
  magnetoelectric and piezomagnetic crystals},\ }\href
  {https://doi.org/10.1080/00150199708222184} {\bibfield  {journal} {\bibinfo
  {journal} {Ferroelectrics}\ }\textbf {\bibinfo {volume} {204}},\ \bibinfo
  {pages} {1} (\bibinfo {year} {1997})}\BibitemShut {NoStop}%
\bibitem [{\citenamefont {W\"ornle}\ \emph {et~al.}(2021)\citenamefont
  {W\"ornle}, \citenamefont {Welter}, \citenamefont {Giraldo}, \citenamefont
  {Lottermoser}, \citenamefont {Fiebig}, \citenamefont {Gambardella},\ and\
  \citenamefont {Degen}}]{Woernle21}%
  \BibitemOpen
  \bibfield  {author} {\bibinfo {author} {\bibfnamefont {M.~S.}\ \bibnamefont
  {W\"ornle}}, \bibinfo {author} {\bibfnamefont {P.}~\bibnamefont {Welter}},
  \bibinfo {author} {\bibfnamefont {M.}~\bibnamefont {Giraldo}}, \bibinfo
  {author} {\bibfnamefont {T.}~\bibnamefont {Lottermoser}}, \bibinfo {author}
  {\bibfnamefont {M.}~\bibnamefont {Fiebig}}, \bibinfo {author} {\bibfnamefont
  {P.}~\bibnamefont {Gambardella}},\ and\ \bibinfo {author} {\bibfnamefont
  {C.~L.}\ \bibnamefont {Degen}},\ }\bibfield  {title} {\bibinfo {title}
  {Coexistence of {B}loch and {N}\'eel walls in a collinear antiferromagnet},\
  }\href {https://doi.org/10.1103/PhysRevB.103.094426} {\bibfield  {journal}
  {\bibinfo  {journal} {Physical Review B}\ }\textbf {\bibinfo {volume}
  {103}},\ \bibinfo {pages} {094426} (\bibinfo {year} {2021})}\BibitemShut
  {NoStop}%
\bibitem [{\citenamefont {Gomonay}\ \emph {et~al.}(2018)\citenamefont
  {Gomonay}, \citenamefont {Baltz}, \citenamefont {Brataas},\ and\
  \citenamefont {Tserkovnyak}}]{Gomonay18}%
  \BibitemOpen
  \bibfield  {author} {\bibinfo {author} {\bibfnamefont {O.}~\bibnamefont
  {Gomonay}}, \bibinfo {author} {\bibfnamefont {V.}~\bibnamefont {Baltz}},
  \bibinfo {author} {\bibfnamefont {A.}~\bibnamefont {Brataas}},\ and\ \bibinfo
  {author} {\bibfnamefont {Y.}~\bibnamefont {Tserkovnyak}},\ }\bibfield
  {title} {\bibinfo {title} {Antiferromagnetic spin textures and dynamics},\
  }\href {https://doi.org/10.1038/s41567-018-0049-4} {\bibfield  {journal}
  {\bibinfo  {journal} {Nature Physics}\ }\textbf {\bibinfo {volume} {14}},\
  \bibinfo {pages} {213} (\bibinfo {year} {2018})}\BibitemShut {NoStop}%
\bibitem [{\citenamefont {Oh}\ \emph {et~al.}(2014)\citenamefont {Oh},
  \citenamefont {Artyukhin}, \citenamefont {S.}, \citenamefont {Yang},\ and\
  \citenamefont {et~al.}}]{Oh14}%
  \BibitemOpen
  \bibfield  {author} {\bibinfo {author} {\bibfnamefont {Y.}~\bibnamefont
  {Oh}}, \bibinfo {author} {\bibnamefont {Artyukhin}}, \bibinfo {author}
  {\bibnamefont {S.}}, \bibinfo {author} {\bibfnamefont {J.}~\bibnamefont
  {Yang}},\ and\ \bibinfo {author} {\bibnamefont {et~al.}},\ }\bibfield
  {title} {\bibinfo {title} {Non-hysteretic colossal magnetoelectricity in a
  collinear antiferromagnet},\ }\bibfield  {journal} {\bibinfo  {journal} {Nat
  Commun}\ }\textbf {\bibinfo {volume} {5}},\ \href
  {https://doi.org/https://doi.org/10.1038/ncomms4201}
  {https://doi.org/10.1038/ncomms4201} (\bibinfo {year} {2014})\BibitemShut
  {NoStop}%
\bibitem [{\citenamefont {Lebrun}\ \emph {et~al.}(2018)\citenamefont {Lebrun},
  \citenamefont {Ross}, \citenamefont {Bender}, \citenamefont {Qaiumzadeh},
  \citenamefont {Baldrati}, \citenamefont {Cramer}, \citenamefont {Brataas},
  \citenamefont {Duine},\ and\ \citenamefont {Kl{\"{a}}ui}}]{Lebrun18}%
  \BibitemOpen
  \bibfield  {author} {\bibinfo {author} {\bibfnamefont {R.}~\bibnamefont
  {Lebrun}}, \bibinfo {author} {\bibfnamefont {A.}~\bibnamefont {Ross}},
  \bibinfo {author} {\bibfnamefont {S.~A.}\ \bibnamefont {Bender}}, \bibinfo
  {author} {\bibfnamefont {A.}~\bibnamefont {Qaiumzadeh}}, \bibinfo {author}
  {\bibfnamefont {L.}~\bibnamefont {Baldrati}}, \bibinfo {author}
  {\bibfnamefont {J.}~\bibnamefont {Cramer}}, \bibinfo {author} {\bibfnamefont
  {A.}~\bibnamefont {Brataas}}, \bibinfo {author} {\bibfnamefont {R.~A.}\
  \bibnamefont {Duine}},\ and\ \bibinfo {author} {\bibfnamefont
  {M.}~\bibnamefont {Kl{\"{a}}ui}},\ }\bibfield  {title} {\bibinfo {title}
  {Tunable long-distance spin transport in a crystalline antiferromagnetic iron
  oxide},\ }\href {https://doi.org/10.1038/s41586-018-0490-7} {\bibfield
  {journal} {\bibinfo  {journal} {Nature}\ }\textbf {\bibinfo {volume} {561}},\
  \bibinfo {pages} {222} (\bibinfo {year} {2018})}\BibitemShut {NoStop}%
\bibitem [{\citenamefont {Bessarab}\ \emph {et~al.}(2019)\citenamefont
  {Bessarab}, \citenamefont {Yudin}, \citenamefont {Gulevich}, \citenamefont
  {Wadley}, \citenamefont {Titov},\ and\ \citenamefont
  {Tretiakov}}]{Bessarab19}%
  \BibitemOpen
  \bibfield  {author} {\bibinfo {author} {\bibfnamefont {P.~F.}\ \bibnamefont
  {Bessarab}}, \bibinfo {author} {\bibfnamefont {D.}~\bibnamefont {Yudin}},
  \bibinfo {author} {\bibfnamefont {D.~R.}\ \bibnamefont {Gulevich}}, \bibinfo
  {author} {\bibfnamefont {P.}~\bibnamefont {Wadley}}, \bibinfo {author}
  {\bibfnamefont {M.}~\bibnamefont {Titov}},\ and\ \bibinfo {author}
  {\bibfnamefont {O.~A.}\ \bibnamefont {Tretiakov}},\ }\bibfield  {title}
  {\bibinfo {title} {Stability and lifetime of antiferromagnetic skyrmions},\
  }\href {https://doi.org/10.1103/PhysRevB.99.140411} {\bibfield  {journal}
  {\bibinfo  {journal} {Phys. Rev. B}\ }\textbf {\bibinfo {volume} {99}},\
  \bibinfo {pages} {140411} (\bibinfo {year} {2019})}\BibitemShut {NoStop}%
\bibitem [{\citenamefont {Upadhyay}\ and\ \citenamefont
  {Sampathkumaran}(2018)}]{Upadhyay18}%
  \BibitemOpen
  \bibfield  {author} {\bibinfo {author} {\bibfnamefont {S.~K.}\ \bibnamefont
  {Upadhyay}}\ and\ \bibinfo {author} {\bibfnamefont {E.~V.}\ \bibnamefont
  {Sampathkumaran}},\ }\bibfield  {title} {\bibinfo {title} {Multiferroicity in
  a spin-chain compound, {Tb}$_2${Ba}{Co}{O}$_5$, with exceptionally large
  magnetodielectric coupling in polycrystalline form},\ }\href
  {https://doi.org/10.1063/1.5037776} {\bibfield  {journal} {\bibinfo
  {journal} {Applied Physics Letters}\ }\textbf {\bibinfo {volume} {112}},\
  \bibinfo {pages} {262902} (\bibinfo {year} {2018})},\ \Eprint
  {https://arxiv.org/abs/https://doi.org/10.1063/1.5037776}
  {https://doi.org/10.1063/1.5037776} \BibitemShut {NoStop}%
\bibitem [{\citenamefont {Peters}\ \emph {et~al.}(2010)\citenamefont {Peters},
  \citenamefont {McCulloch},\ and\ \citenamefont {Selke}}]{Peters10}%
  \BibitemOpen
  \bibfield  {author} {\bibinfo {author} {\bibfnamefont {D.}~\bibnamefont
  {Peters}}, \bibinfo {author} {\bibfnamefont {I.~P.}\ \bibnamefont
  {McCulloch}},\ and\ \bibinfo {author} {\bibfnamefont {W.}~\bibnamefont
  {Selke}},\ }\bibfield  {title} {\bibinfo {title} {Quantum heisenberg
  antiferromagnetic chains with exchange and single-ion anisotropies},\ }\href
  {https://doi.org/10.1088/1742-6596/200/2/022046} {\bibfield  {journal}
  {\bibinfo  {journal} {Journal of Physics: Conference Series}\ }\textbf
  {\bibinfo {volume} {200}},\ \bibinfo {pages} {022046} (\bibinfo {year}
  {2010})}\BibitemShut {NoStop}%
\bibitem [{\citenamefont {Prokhnenko}\ \emph {et~al.}(2021)\citenamefont
  {Prokhnenko}, \citenamefont {Marmorini}, \citenamefont {Nikitin},
  \citenamefont {Yamamoto}, \citenamefont {Gazizulina}, \citenamefont
  {Bartkowiak}, \citenamefont {Ponomaryov}, \citenamefont {Zvyagin},
  \citenamefont {Nojiri}, \citenamefont {D\'{\i}az-Ortega}, \citenamefont
  {Anovitz}, \citenamefont {Kolesnikov},\ and\ \citenamefont
  {Podlesnyak}}]{Prokhnenko21}%
  \BibitemOpen
  \bibfield  {author} {\bibinfo {author} {\bibfnamefont {O.}~\bibnamefont
  {Prokhnenko}}, \bibinfo {author} {\bibfnamefont {G.}~\bibnamefont
  {Marmorini}}, \bibinfo {author} {\bibfnamefont {S.~E.}\ \bibnamefont
  {Nikitin}}, \bibinfo {author} {\bibfnamefont {D.}~\bibnamefont {Yamamoto}},
  \bibinfo {author} {\bibfnamefont {A.}~\bibnamefont {Gazizulina}}, \bibinfo
  {author} {\bibfnamefont {M.}~\bibnamefont {Bartkowiak}}, \bibinfo {author}
  {\bibfnamefont {A.~N.}\ \bibnamefont {Ponomaryov}}, \bibinfo {author}
  {\bibfnamefont {S.~A.}\ \bibnamefont {Zvyagin}}, \bibinfo {author}
  {\bibfnamefont {H.}~\bibnamefont {Nojiri}}, \bibinfo {author} {\bibfnamefont
  {I.~F.}\ \bibnamefont {D\'{\i}az-Ortega}}, \bibinfo {author} {\bibfnamefont
  {L.~M.}\ \bibnamefont {Anovitz}}, \bibinfo {author} {\bibfnamefont {A.~I.}\
  \bibnamefont {Kolesnikov}},\ and\ \bibinfo {author} {\bibfnamefont
  {A.}~\bibnamefont {Podlesnyak}},\ }\bibfield  {title} {\bibinfo {title}
  {High-field spin-flop state in green dioptase},\ }\href
  {https://doi.org/10.1103/PhysRevB.103.014427} {\bibfield  {journal} {\bibinfo
   {journal} {Phys. Rev. B}\ }\textbf {\bibinfo {volume} {103}},\ \bibinfo
  {pages} {014427} (\bibinfo {year} {2021})}\BibitemShut {NoStop}%
\bibitem [{\citenamefont {Fischer}\ \emph {et~al.}(2020)\citenamefont
  {Fischer}, \citenamefont {Sanz-Hern{\'{a}}ndez}, \citenamefont {Streubel},\
  and\ \citenamefont {Fern{\'{a}}ndez-Pacheco}}]{Fischer20}%
  \BibitemOpen
  \bibfield  {author} {\bibinfo {author} {\bibfnamefont {P.}~\bibnamefont
  {Fischer}}, \bibinfo {author} {\bibfnamefont {D.}~\bibnamefont
  {Sanz-Hern{\'{a}}ndez}}, \bibinfo {author} {\bibfnamefont {R.}~\bibnamefont
  {Streubel}},\ and\ \bibinfo {author} {\bibfnamefont {A.}~\bibnamefont
  {Fern{\'{a}}ndez-Pacheco}},\ }\bibfield  {title} {\bibinfo {title} {Launching
  a new dimension with 3{D} magnetic nanostructures},\ }\href
  {https://doi.org/10.1063/1.5134474} {\bibfield  {journal} {\bibinfo
  {journal} {{APL} Materials}\ }\textbf {\bibinfo {volume} {8}},\ \bibinfo
  {pages} {010701} (\bibinfo {year} {2020})}\BibitemShut {NoStop}%
\bibitem [{\citenamefont {Sheka}(2021)}]{Sheka21b}%
  \BibitemOpen
  \bibfield  {author} {\bibinfo {author} {\bibfnamefont {D.~D.}\ \bibnamefont
  {Sheka}},\ }\bibfield  {title} {\bibinfo {title} {A perspective on
  curvilinear magnetism},\ }\href {https://doi.org/10.1063/5.0048891}
  {\bibfield  {journal} {\bibinfo  {journal} {Applied Physics Letters}\
  }\textbf {\bibinfo {volume} {118}},\ \bibinfo {pages} {230502} (\bibinfo
  {year} {2021})}\BibitemShut {NoStop}%
\bibitem [{\citenamefont {Streubel}\ \emph {et~al.}(2021)\citenamefont
  {Streubel}, \citenamefont {Tsymbal},\ and\ \citenamefont
  {Fischer}}]{Streubel21}%
  \BibitemOpen
  \bibfield  {author} {\bibinfo {author} {\bibfnamefont {R.}~\bibnamefont
  {Streubel}}, \bibinfo {author} {\bibfnamefont {E.~Y.}\ \bibnamefont
  {Tsymbal}},\ and\ \bibinfo {author} {\bibfnamefont {P.}~\bibnamefont
  {Fischer}},\ }\bibfield  {title} {\bibinfo {title} {Magnetism in curved
  geometries},\ }\href {https://doi.org/10.1063/5.0054025} {\bibfield
  {journal} {\bibinfo  {journal} {Journal of Applied Physics}\ }\textbf
  {\bibinfo {volume} {129}},\ \bibinfo {pages} {210902} (\bibinfo {year}
  {2021})}\BibitemShut {NoStop}%
\bibitem [{\citenamefont {Makarov}\ \emph {et~al.}(2022)\citenamefont
  {Makarov}, \citenamefont {Volkov}, \citenamefont {K\'{a}kay}, \citenamefont
  {Pylypovskyi}, \citenamefont {Budinsk\'{a}},\ and\ \citenamefont
  {Dobrovolskiy}}]{Makarov22}%
  \BibitemOpen
  \bibfield  {author} {\bibinfo {author} {\bibfnamefont {D.}~\bibnamefont
  {Makarov}}, \bibinfo {author} {\bibfnamefont {O.~M.}\ \bibnamefont {Volkov}},
  \bibinfo {author} {\bibfnamefont {A.}~\bibnamefont {K\'{a}kay}}, \bibinfo
  {author} {\bibfnamefont {O.~V.}\ \bibnamefont {Pylypovskyi}}, \bibinfo
  {author} {\bibfnamefont {B.}~\bibnamefont {Budinsk\'{a}}},\ and\ \bibinfo
  {author} {\bibfnamefont {O.~V.}\ \bibnamefont {Dobrovolskiy}},\ }\bibfield
  {title} {\bibinfo {title} {New dimension in magnetism and superconductivity:
  {3D} and curvilinear nano-architectures},\ }\href
  {https://doi.org/10.1002/adma.202101758} {\bibfield  {journal} {\bibinfo
  {journal} {Advanced Materials}\ }\textbf {\bibinfo {volume} {34}},\ \bibinfo
  {pages} {2101758} (\bibinfo {year} {2022})}\BibitemShut {NoStop}%
\bibitem [{\citenamefont {Pylypovskyi}\ \emph {et~al.}(2020)\citenamefont
  {Pylypovskyi}, \citenamefont {Kononenko}, \citenamefont {Yershov},
  \citenamefont {R{\"{o}}{\ss}ler}, \citenamefont {Tomilo}, \citenamefont
  {Fassbender}, \citenamefont {van~den Brink}, \citenamefont {Makarov},\ and\
  \citenamefont {Sheka}}]{Pylypovskyi20}%
  \BibitemOpen
  \bibfield  {author} {\bibinfo {author} {\bibfnamefont {O.~V.}\ \bibnamefont
  {Pylypovskyi}}, \bibinfo {author} {\bibfnamefont {D.~Y.}\ \bibnamefont
  {Kononenko}}, \bibinfo {author} {\bibfnamefont {K.~V.}\ \bibnamefont
  {Yershov}}, \bibinfo {author} {\bibfnamefont {U.~K.}\ \bibnamefont
  {R{\"{o}}{\ss}ler}}, \bibinfo {author} {\bibfnamefont {A.~V.}\ \bibnamefont
  {Tomilo}}, \bibinfo {author} {\bibfnamefont {J.}~\bibnamefont {Fassbender}},
  \bibinfo {author} {\bibfnamefont {J.}~\bibnamefont {van~den Brink}}, \bibinfo
  {author} {\bibfnamefont {D.}~\bibnamefont {Makarov}},\ and\ \bibinfo {author}
  {\bibfnamefont {D.~D.}\ \bibnamefont {Sheka}},\ }\bibfield  {title} {\bibinfo
  {title} {Curvilinear one-dimensional antiferromagnets},\ }\href
  {https://doi.org/10.1021/acs.nanolett.0c03246} {\bibfield  {journal}
  {\bibinfo  {journal} {Nano Letters}\ }\textbf {\bibinfo {volume} {20}},\
  \bibinfo {pages} {8157} (\bibinfo {year} {2020})}\BibitemShut {NoStop}%
\bibitem [{\citenamefont {Wu}\ and\ \citenamefont {Lan}(2022)}]{Wu22}%
  \BibitemOpen
  \bibfield  {author} {\bibinfo {author} {\bibfnamefont {H.}~\bibnamefont
  {Wu}}\ and\ \bibinfo {author} {\bibfnamefont {J.}~\bibnamefont {Lan}},\
  }\bibfield  {title} {\bibinfo {title} {Curvilinear manipulation of polarized
  spin waves},\ }\href {https://doi.org/10.1103/PhysRevB.105.174427} {\bibfield
   {journal} {\bibinfo  {journal} {Phys. Rev. B}\ }\textbf {\bibinfo {volume}
  {105}},\ \bibinfo {pages} {174427} (\bibinfo {year} {2022})}\BibitemShut
  {NoStop}%
\bibitem [{\citenamefont {Castillo-Sep{\'{u}}lveda}\ \emph
  {et~al.}(2017)\citenamefont {Castillo-Sep{\'{u}}lveda}, \citenamefont
  {Escobar}, \citenamefont {Altbir}, \citenamefont {Krizanac},\ and\
  \citenamefont {Vedmedenko}}]{Castillo-Sepulveda17}%
  \BibitemOpen
  \bibfield  {author} {\bibinfo {author} {\bibfnamefont {S.}~\bibnamefont
  {Castillo-Sep{\'{u}}lveda}}, \bibinfo {author} {\bibfnamefont {R.~A.}\
  \bibnamefont {Escobar}}, \bibinfo {author} {\bibfnamefont {D.}~\bibnamefont
  {Altbir}}, \bibinfo {author} {\bibfnamefont {M.}~\bibnamefont {Krizanac}},\
  and\ \bibinfo {author} {\bibfnamefont {E.~Y.}\ \bibnamefont {Vedmedenko}},\
  }\bibfield  {title} {\bibinfo {title} {Magnetic {M}{\"{o}}bius stripe without
  frustration: Noncollinear metastable states},\ }\href
  {https://doi.org/10.1103/physrevb.96.024426} {\bibfield  {journal} {\bibinfo
  {journal} {Physical Review B}\ }\textbf {\bibinfo {volume} {96}},\ \bibinfo
  {pages} {024426} (\bibinfo {year} {2017})}\BibitemShut {NoStop}%
\bibitem [{\citenamefont {Yershov}(2022)}]{Yershov22a}%
  \BibitemOpen
  \bibfield  {author} {\bibinfo {author} {\bibfnamefont {K.~V.}\ \bibnamefont
  {Yershov}},\ }\bibfield  {title} {\bibinfo {title} {Dynamics of domain walls
  in curved antiferromagnetic wires},\ }\href
  {https://doi.org/10.1103/PhysRevB.105.064407} {\bibfield  {journal} {\bibinfo
   {journal} {Phys. Rev. B}\ }\textbf {\bibinfo {volume} {105}},\ \bibinfo
  {pages} {064407} (\bibinfo {year} {2022})}\BibitemShut {NoStop}%
\bibitem [{\citenamefont {Yershov}\ \emph {et~al.}(2022)\citenamefont
  {Yershov}, \citenamefont {K{\'{a}}kay},\ and\ \citenamefont
  {Kravchuk}}]{Yershov22}%
  \BibitemOpen
  \bibfield  {author} {\bibinfo {author} {\bibfnamefont {K.~V.}\ \bibnamefont
  {Yershov}}, \bibinfo {author} {\bibfnamefont {A.}~\bibnamefont
  {K{\'{a}}kay}},\ and\ \bibinfo {author} {\bibfnamefont {V.~P.}\ \bibnamefont
  {Kravchuk}},\ }\bibfield  {title} {\bibinfo {title} {Curvature-induced drift
  and deformation of magnetic skyrmions: Comparison of the ferromagnetic and
  antiferromagnetic cases},\ }\href
  {https://doi.org/10.1103/physrevb.105.054425} {\bibfield  {journal} {\bibinfo
   {journal} {Physical Review B}\ }\textbf {\bibinfo {volume} {105}},\ \bibinfo
  {pages} {054425} (\bibinfo {year} {2022})}\BibitemShut {NoStop}%
\bibitem [{\citenamefont {Pylypovskyi}\ \emph {et~al.}(2021)\citenamefont
  {Pylypovskyi}, \citenamefont {Borysenko}, \citenamefont {Fassbender},
  \citenamefont {Sheka},\ and\ \citenamefont {Makarov}}]{Pylypovskyi21f}%
  \BibitemOpen
  \bibfield  {author} {\bibinfo {author} {\bibfnamefont {O.~V.}\ \bibnamefont
  {Pylypovskyi}}, \bibinfo {author} {\bibfnamefont {Y.~A.}\ \bibnamefont
  {Borysenko}}, \bibinfo {author} {\bibfnamefont {J.}~\bibnamefont
  {Fassbender}}, \bibinfo {author} {\bibfnamefont {D.~D.}\ \bibnamefont
  {Sheka}},\ and\ \bibinfo {author} {\bibfnamefont {D.}~\bibnamefont
  {Makarov}},\ }\bibfield  {title} {\bibinfo {title} {Curvature-driven
  homogeneous {D}zyaloshinskii--{M}oriya interaction and emergent weak
  ferromagnetism in anisotropic antiferromagnetic spin chains},\ }\href
  {https://doi.org/10.1063/5.0048823} {\bibfield  {journal} {\bibinfo
  {journal} {Applied Physics Letters}\ }\textbf {\bibinfo {volume} {118}},\
  \bibinfo {pages} {182405} (\bibinfo {year} {2021})}\BibitemShut {NoStop}%
\bibitem [{\citenamefont {Papanicolaou}\ and\ \citenamefont
  {Zakrzewski}(1995)}]{Papanicolaou95}%
  \BibitemOpen
  \bibfield  {author} {\bibinfo {author} {\bibfnamefont {N.}~\bibnamefont
  {Papanicolaou}}\ and\ \bibinfo {author} {\bibfnamefont {W.~J.}\ \bibnamefont
  {Zakrzewski}},\ }\bibfield  {title} {\bibinfo {title} {Dynamics of
  interacting magnetic vortices in a model {Landau--Lifshitz} equation},\
  }\href
  {http://www.sciencedirect.com/science/article/B6TVK-4031TPX-7/2/e8a698e741096bb98869389ad030fe69}
  {\bibfield  {journal} {\bibinfo  {journal} {Physica D: Nonlinear Phenomena}\
  }\textbf {\bibinfo {volume} {80}},\ \bibinfo {pages} {225} (\bibinfo {year}
  {1995})}\BibitemShut {NoStop}%
\bibitem [{\citenamefont {Olver}\ \emph {et~al.}(2010)\citenamefont {Olver},
  \citenamefont {Lozier}, \citenamefont {Boisvert},\ and\ \citenamefont
  {Clark}}]{NIST10}%
  \BibitemOpen
  \bibinfo {editor} {\bibfnamefont {F.~W.~J.}\ \bibnamefont {Olver}}, \bibinfo
  {editor} {\bibfnamefont {D.~W.}\ \bibnamefont {Lozier}}, \bibinfo {editor}
  {\bibfnamefont {R.~F.}\ \bibnamefont {Boisvert}},\ and\ \bibinfo {editor}
  {\bibfnamefont {C.~W.}\ \bibnamefont {Clark}},\ eds.,\ \href
  {http://www.cambridge.org/us/academic/subjects/mathematics/abstract-analysis/nist-handbook-mathematical-functions}
  {\emph {\bibinfo {title} {NIST Handbook of Mathematical Functions}}}\
  (\bibinfo  {publisher} {Cambridge University Press},\ \bibinfo {address} {New
  York, NY},\ \bibinfo {year} {2010})\BibitemShut {NoStop}%
\bibitem [{\citenamefont {Sheka}\ \emph {et~al.}(2015)\citenamefont {Sheka},
  \citenamefont {Kravchuk},\ and\ \citenamefont {Gaididei}}]{Sheka15}%
  \BibitemOpen
  \bibfield  {author} {\bibinfo {author} {\bibfnamefont {D.~D.}\ \bibnamefont
  {Sheka}}, \bibinfo {author} {\bibfnamefont {V.~P.}\ \bibnamefont
  {Kravchuk}},\ and\ \bibinfo {author} {\bibfnamefont {Y.}~\bibnamefont
  {Gaididei}},\ }\bibfield  {title} {\bibinfo {title} {Curvature effects in
  statics and dynamics of low dimensional magnets},\ }\href
  {https://doi.org/10.1088/1751-8113/48/12/125202} {\bibfield  {journal}
  {\bibinfo  {journal} {Journal of Physics A: Mathematical and Theoretical}\
  }\textbf {\bibinfo {volume} {48}},\ \bibinfo {pages} {125202} (\bibinfo
  {year} {2015})}\BibitemShut {NoStop}%
\bibitem [{\citenamefont {Bogdanov}\ \emph {et~al.}(2007)\citenamefont
  {Bogdanov}, \citenamefont {Zhuravlev},\ and\ \citenamefont
  {R\"{o}{\ss}ler}}]{Bogdanov07}%
  \BibitemOpen
  \bibfield  {author} {\bibinfo {author} {\bibfnamefont {A.~N.}\ \bibnamefont
  {Bogdanov}}, \bibinfo {author} {\bibfnamefont {A.~V.}\ \bibnamefont
  {Zhuravlev}},\ and\ \bibinfo {author} {\bibfnamefont {U.~K.}\ \bibnamefont
  {R\"{o}{\ss}ler}},\ }\bibfield  {title} {\bibinfo {title} {Spin-flop
  transition in uniaxial antiferromagnets: Magnetic phases, reorientation
  effects, and multidomain states},\ }\href
  {https://doi.org/10.1103/physrevb.75.094425} {\bibfield  {journal} {\bibinfo
  {journal} {Physical Review B}\ }\textbf {\bibinfo {volume} {75}},\ \bibinfo
  {pages} {094425} (\bibinfo {year} {2007})}\BibitemShut {NoStop}%
\bibitem [{\citenamefont {Medvedovskaya}\ and\ \citenamefont
  {Chepurnykh}(1985)}]{Medvedovskaya85}%
  \BibitemOpen
  \bibfield  {author} {\bibinfo {author} {\bibfnamefont {O.~G.}\ \bibnamefont
  {Medvedovskaya}}\ and\ \bibinfo {author} {\bibfnamefont {G.~K.}\ \bibnamefont
  {Chepurnykh}},\ }\bibfield  {title} {\bibinfo {title} {Dzyaloshinskii
  interaction effect on orienational phase-transitions in antiferromagnets},\
  }\href
  {http://www.mathnet.ru/php/archive.phtml?wshow=paper&jrnid=ftt&paperid=1824}
  {\bibfield  {journal} {\bibinfo  {journal} {Solid state physics (Soviet
  Fizika tverdogo tela)}\ }\textbf {\bibinfo {volume} {27}},\ \bibinfo {pages}
  {718} (\bibinfo {year} {1985})}\BibitemShut {NoStop}%
\bibitem [{\citenamefont {Thio}\ \emph {et~al.}(1990)\citenamefont {Thio},
  \citenamefont {Chen}, \citenamefont {Freer}, \citenamefont {Gabbe},
  \citenamefont {Jenssen}, \citenamefont {Kastner}, \citenamefont {Picone},
  \citenamefont {Preyer},\ and\ \citenamefont {Birgeneau}}]{Thio90}%
  \BibitemOpen
  \bibfield  {author} {\bibinfo {author} {\bibfnamefont {T.}~\bibnamefont
  {Thio}}, \bibinfo {author} {\bibfnamefont {C.~Y.}\ \bibnamefont {Chen}},
  \bibinfo {author} {\bibfnamefont {B.~S.}\ \bibnamefont {Freer}}, \bibinfo
  {author} {\bibfnamefont {D.~R.}\ \bibnamefont {Gabbe}}, \bibinfo {author}
  {\bibfnamefont {H.~P.}\ \bibnamefont {Jenssen}}, \bibinfo {author}
  {\bibfnamefont {M.~A.}\ \bibnamefont {Kastner}}, \bibinfo {author}
  {\bibfnamefont {P.~J.}\ \bibnamefont {Picone}}, \bibinfo {author}
  {\bibfnamefont {N.~W.}\ \bibnamefont {Preyer}},\ and\ \bibinfo {author}
  {\bibfnamefont {R.~J.}\ \bibnamefont {Birgeneau}},\ }\bibfield  {title}
  {\bibinfo {title} {Magnetoresistance and the spin-flop transition in
  single-crystal $\text{La}_{2}\text{CuO}_{4+\text{y}}$},\ }\href
  {https://doi.org/10.1103/PhysRevB.41.231} {\bibfield  {journal} {\bibinfo
  {journal} {Phys. Rev. B}\ }\textbf {\bibinfo {volume} {41}},\ \bibinfo
  {pages} {231} (\bibinfo {year} {1990})}\BibitemShut {NoStop}%
\bibitem [{\citenamefont {Tsukada}\ \emph {et~al.}(2001)\citenamefont
  {Tsukada}, \citenamefont {Takeya}, \citenamefont {Masuda},\ and\
  \citenamefont {Uchinokura}}]{Tsukada01}%
  \BibitemOpen
  \bibfield  {author} {\bibinfo {author} {\bibfnamefont {I.}~\bibnamefont
  {Tsukada}}, \bibinfo {author} {\bibfnamefont {J.}~\bibnamefont {Takeya}},
  \bibinfo {author} {\bibfnamefont {T.}~\bibnamefont {Masuda}},\ and\ \bibinfo
  {author} {\bibfnamefont {K.}~\bibnamefont {Uchinokura}},\ }\bibfield  {title}
  {\bibinfo {title} {Two-stage spin-flop transitions in the antiferromagnetic
  spin chain $\text{BaCu}_{2}\text{Si}_{2}\text{O}_{7}$},\ }\href
  {https://doi.org/10.1103/PhysRevLett.87.127203} {\bibfield  {journal}
  {\bibinfo  {journal} {Phys. Rev. Lett.}\ }\textbf {\bibinfo {volume} {87}},\
  \bibinfo {pages} {127203} (\bibinfo {year} {2001})}\BibitemShut {NoStop}%
\bibitem [{\citenamefont {Sorolla}\ \emph {et~al.}(2019)\citenamefont
  {Sorolla}, \citenamefont {Wang}, \citenamefont {Makarenko},\ and\
  \citenamefont {Jacobson}}]{Sorolla19}%
  \BibitemOpen
  \bibfield  {author} {\bibinfo {author} {\bibfnamefont {M.~G.}\ \bibnamefont
  {Sorolla}}, \bibinfo {author} {\bibfnamefont {X.}~\bibnamefont {Wang}},
  \bibinfo {author} {\bibfnamefont {T.}~\bibnamefont {Makarenko}},\ and\
  \bibinfo {author} {\bibfnamefont {A.~J.}\ \bibnamefont {Jacobson}},\
  }\bibfield  {title} {\bibinfo {title} {A large spin, magnetically
  anisotropic, octanuclear vanadium(iii) wheel},\ }\href
  {https://doi.org/10.1039/c8cc08372j} {\bibfield  {journal} {\bibinfo
  {journal} {Chemical Communications}\ }\textbf {\bibinfo {volume} {55}},\
  \bibinfo {pages} {342} (\bibinfo {year} {2019})}\BibitemShut {NoStop}%
\bibitem [{\citenamefont {Fu}\ \emph {et~al.}(2021)\citenamefont {Fu},
  \citenamefont {Wei}, \citenamefont {Niu},\ and\ \citenamefont {Wang}}]{Fu21}%
  \BibitemOpen
  \bibfield  {author} {\bibinfo {author} {\bibfnamefont {X.-X.}\ \bibnamefont
  {Fu}}, \bibinfo {author} {\bibfnamefont {F.}~\bibnamefont {Wei}}, \bibinfo
  {author} {\bibfnamefont {Y.}~\bibnamefont {Niu}},\ and\ \bibinfo {author}
  {\bibfnamefont {C.-K.}\ \bibnamefont {Wang}},\ }\bibfield  {title} {\bibinfo
  {title} {Designing high-performance spin filters and valves based on
  metal-salophen molecular chains},\ }\href
  {https://doi.org/https://doi.org/10.1016/j.physe.2021.114737} {\bibfield
  {journal} {\bibinfo  {journal} {Physica E: Low-dimensional Systems and
  Nanostructures}\ }\textbf {\bibinfo {volume} {131}},\ \bibinfo {pages}
  {114737} (\bibinfo {year} {2021})}\BibitemShut {NoStop}%
\bibitem [{\citenamefont {Mizoguchi}\ \emph {et~al.}(2007)\citenamefont
  {Mizoguchi}, \citenamefont {Tanaka}, \citenamefont {Ojima}, \citenamefont
  {Sano}, \citenamefont {Nagatori}, \citenamefont {Sakamoto}, \citenamefont
  {Yonezawa}, \citenamefont {Aoki}, \citenamefont {Sato}, \citenamefont
  {Furukawa},\ and\ \citenamefont {Nakamura}}]{Mizoguchi07}%
  \BibitemOpen
  \bibfield  {author} {\bibinfo {author} {\bibfnamefont {K.}~\bibnamefont
  {Mizoguchi}}, \bibinfo {author} {\bibfnamefont {S.}~\bibnamefont {Tanaka}},
  \bibinfo {author} {\bibfnamefont {M.}~\bibnamefont {Ojima}}, \bibinfo
  {author} {\bibfnamefont {S.}~\bibnamefont {Sano}}, \bibinfo {author}
  {\bibfnamefont {M.}~\bibnamefont {Nagatori}}, \bibinfo {author}
  {\bibfnamefont {H.}~\bibnamefont {Sakamoto}}, \bibinfo {author}
  {\bibfnamefont {Y.}~\bibnamefont {Yonezawa}}, \bibinfo {author}
  {\bibfnamefont {Y.}~\bibnamefont {Aoki}}, \bibinfo {author} {\bibfnamefont
  {H.}~\bibnamefont {Sato}}, \bibinfo {author} {\bibfnamefont {K.}~\bibnamefont
  {Furukawa}},\ and\ \bibinfo {author} {\bibfnamefont {T.}~\bibnamefont
  {Nakamura}},\ }\bibfield  {title} {\bibinfo {title} {{AF}-like ground state
  of {Mn}-{DNA} and charge transfer from {Fe} to base-$\pi$-band in
  {Fe}-{DNA}},\ }\href {https://doi.org/10.1143/jpsj.76.043801} {\bibfield
  {journal} {\bibinfo  {journal} {Journal of the Physical Society of Japan}\
  }\textbf {\bibinfo {volume} {76}},\ \bibinfo {pages} {043801} (\bibinfo
  {year} {2007})}\BibitemShut {NoStop}%
\bibitem [{\citenamefont {Kesama}\ \emph {et~al.}(2018)\citenamefont {Kesama},
  \citenamefont {Yun}, \citenamefont {Dugasani}, \citenamefont {Jung},\ and\
  \citenamefont {Park}}]{Kesama18}%
  \BibitemOpen
  \bibfield  {author} {\bibinfo {author} {\bibfnamefont {M.~R.}\ \bibnamefont
  {Kesama}}, \bibinfo {author} {\bibfnamefont {B.~K.}\ \bibnamefont {Yun}},
  \bibinfo {author} {\bibfnamefont {S.~R.}\ \bibnamefont {Dugasani}}, \bibinfo
  {author} {\bibfnamefont {J.~H.}\ \bibnamefont {Jung}},\ and\ \bibinfo
  {author} {\bibfnamefont {S.~H.}\ \bibnamefont {Park}},\ }\bibfield  {title}
  {\bibinfo {title} {Enhancing the electrical, optical, and magnetic
  characteristics of {DNA} thin films through {Mn}$^{2+}$ fortification},\
  }\href {https://doi.org/10.1016/j.colsurfb.2018.04.023} {\bibfield  {journal}
  {\bibinfo  {journal} {Colloids and Surfaces B: Biointerfaces}\ }\textbf
  {\bibinfo {volume} {167}},\ \bibinfo {pages} {197} (\bibinfo {year}
  {2018})}\BibitemShut {NoStop}%
\bibitem [{\citenamefont {Loth}\ \emph {et~al.}(2012)\citenamefont {Loth},
  \citenamefont {Baumann}, \citenamefont {Lutz}, \citenamefont {Eigler},\ and\
  \citenamefont {Heinrich}}]{Loth12}%
  \BibitemOpen
  \bibfield  {author} {\bibinfo {author} {\bibfnamefont {S.}~\bibnamefont
  {Loth}}, \bibinfo {author} {\bibfnamefont {S.}~\bibnamefont {Baumann}},
  \bibinfo {author} {\bibfnamefont {C.~P.}\ \bibnamefont {Lutz}}, \bibinfo
  {author} {\bibfnamefont {D.~M.}\ \bibnamefont {Eigler}},\ and\ \bibinfo
  {author} {\bibfnamefont {A.~J.}\ \bibnamefont {Heinrich}},\ }\bibfield
  {title} {\bibinfo {title} {Bistability in atomic-scale antiferromagnets},\
  }\href {https://doi.org/10.1126/science.1214131} {\bibfield  {journal}
  {\bibinfo  {journal} {Science}\ }\textbf {\bibinfo {volume} {335}},\ \bibinfo
  {pages} {196} (\bibinfo {year} {2012})}\BibitemShut {NoStop}%
\bibitem [{\citenamefont {Khajetoorians}\ \emph {et~al.}(2012)\citenamefont
  {Khajetoorians}, \citenamefont {Wiebe}, \citenamefont {Chilian},
  \citenamefont {Lounis}, \citenamefont {Bl\"{u}gel},\ and\ \citenamefont
  {Wiesendanger}}]{Khajetoorians12}%
  \BibitemOpen
  \bibfield  {author} {\bibinfo {author} {\bibfnamefont {A.~A.}\ \bibnamefont
  {Khajetoorians}}, \bibinfo {author} {\bibfnamefont {J.}~\bibnamefont
  {Wiebe}}, \bibinfo {author} {\bibfnamefont {B.}~\bibnamefont {Chilian}},
  \bibinfo {author} {\bibfnamefont {S.}~\bibnamefont {Lounis}}, \bibinfo
  {author} {\bibfnamefont {S.}~\bibnamefont {Bl\"{u}gel}},\ and\ \bibinfo
  {author} {\bibfnamefont {R.}~\bibnamefont {Wiesendanger}},\ }\bibfield
  {title} {\bibinfo {title} {Atom-by-atom engineering and magnetometry of
  tailored nanomagnets},\ }\href {https://doi.org/10.1038/nphys2299} {\bibfield
   {journal} {\bibinfo  {journal} {Nature Physics}\ }\textbf {\bibinfo {volume}
  {8}},\ \bibinfo {pages} {497} (\bibinfo {year} {2012})}\BibitemShut {NoStop}%
\bibitem [{\citenamefont {Santana}\ \emph {et~al.}(2012)\citenamefont
  {Santana}, \citenamefont {Ferreira}, \citenamefont {Sabino}, \citenamefont
  {Carvalho}, \citenamefont {Pe{\~{n}}a},\ and\ \citenamefont
  {Calvo}}]{Santana12}%
  \BibitemOpen
  \bibfield  {author} {\bibinfo {author} {\bibfnamefont {R.~C.}\ \bibnamefont
  {Santana}}, \bibinfo {author} {\bibfnamefont {B.~N.}\ \bibnamefont
  {Ferreira}}, \bibinfo {author} {\bibfnamefont {J.~R.}\ \bibnamefont
  {Sabino}}, \bibinfo {author} {\bibfnamefont {J.~F.}\ \bibnamefont
  {Carvalho}}, \bibinfo {author} {\bibfnamefont {O.}~\bibnamefont
  {Pe{\~{n}}a}},\ and\ \bibinfo {author} {\bibfnamefont {R.}~\bibnamefont
  {Calvo}},\ }\bibfield  {title} {\bibinfo {title} {Structure and magnetism of
  catena-poly[copper({II})-$\mu$-dichloro-{L}-lysine]hemihydrate: Copper chains
  with monochloride bridges},\ }\href
  {https://doi.org/10.1016/j.poly.2012.08.045} {\bibfield  {journal} {\bibinfo
  {journal} {Polyhedron}\ }\textbf {\bibinfo {volume} {47}},\ \bibinfo {pages}
  {53} (\bibinfo {year} {2012})}\BibitemShut {NoStop}%
\bibitem [{\citenamefont {Drahoš~B}(2015)}]{Drahos15}%
  \BibitemOpen
  \bibfield  {author} {\bibinfo {author} {\bibfnamefont {T.~Z.}\ \bibnamefont
  {Drahoš~B}, \bibfnamefont {Herchel~R}},\ }\bibfield  {title} {\bibinfo
  {title} {Structural, magnetic, and redox diversity of first-row transition
  metal complexes of a pyridine-based macrocycle: well-marked trends supported
  by theoretical dft calculations},\ }\href {https://doi.org/10.1021/ic503054m}
  {\bibfield  {journal} {\bibinfo  {journal} {Inorg Chem.}\ }\textbf {\bibinfo
  {volume} {54}},\ \bibinfo {pages} {3352} (\bibinfo {year}
  {2015})}\BibitemShut {NoStop}%
\bibitem [{\citenamefont {Furukawa}\ \emph {et~al.}(2008)\citenamefont
  {Furukawa}, \citenamefont {Kiuchi}, \citenamefont {Kumagai}, \citenamefont
  {Ajiro}, \citenamefont {Narumi}, \citenamefont {Iwaki}, \citenamefont
  {Kindo}, \citenamefont {Bianchi}, \citenamefont {Carretta}, \citenamefont
  {Timco},\ and\ \citenamefont {Winpenny}}]{Furukawa08}%
  \BibitemOpen
  \bibfield  {author} {\bibinfo {author} {\bibfnamefont {Y.}~\bibnamefont
  {Furukawa}}, \bibinfo {author} {\bibfnamefont {K.}~\bibnamefont {Kiuchi}},
  \bibinfo {author} {\bibfnamefont {K.-i.}\ \bibnamefont {Kumagai}}, \bibinfo
  {author} {\bibfnamefont {Y.}~\bibnamefont {Ajiro}}, \bibinfo {author}
  {\bibfnamefont {Y.}~\bibnamefont {Narumi}}, \bibinfo {author} {\bibfnamefont
  {M.}~\bibnamefont {Iwaki}}, \bibinfo {author} {\bibfnamefont
  {K.}~\bibnamefont {Kindo}}, \bibinfo {author} {\bibfnamefont
  {A.}~\bibnamefont {Bianchi}}, \bibinfo {author} {\bibfnamefont
  {S.}~\bibnamefont {Carretta}}, \bibinfo {author} {\bibfnamefont {G.~A.}\
  \bibnamefont {Timco}},\ and\ \bibinfo {author} {\bibfnamefont {R.~E.~P.}\
  \bibnamefont {Winpenny}},\ }\bibfield  {title} {\bibinfo {title} {Topological
  effects on the magnetic properties of closed and open ring-shaped cr-based
  antiferromagnetic nanomagnets},\ }\href
  {https://doi.org/10.1103/PhysRevB.78.092402} {\bibfield  {journal} {\bibinfo
  {journal} {Physical Review B}\ }\textbf {\bibinfo {volume} {78}},\ \bibinfo
  {pages} {092402} (\bibinfo {year} {2008})}\BibitemShut {NoStop}%
\bibitem [{\citenamefont {Guidi}\ \emph {et~al.}(2015)\citenamefont {Guidi},
  \citenamefont {Gillon}, \citenamefont {Mason}, \citenamefont {Garlatti},
  \citenamefont {Carretta}, \citenamefont {Santini}, \citenamefont {Stunault},
  \citenamefont {Caciuffo}, \citenamefont {van Slageren}, \citenamefont
  {Klemke}, \citenamefont {Cousson}, \citenamefont {Timco},\ and\ \citenamefont
  {Winpenny}}]{Guidi15}%
  \BibitemOpen
  \bibfield  {author} {\bibinfo {author} {\bibfnamefont {T.}~\bibnamefont
  {Guidi}}, \bibinfo {author} {\bibfnamefont {B.}~\bibnamefont {Gillon}},
  \bibinfo {author} {\bibfnamefont {S.~A.}\ \bibnamefont {Mason}}, \bibinfo
  {author} {\bibfnamefont {E.}~\bibnamefont {Garlatti}}, \bibinfo {author}
  {\bibfnamefont {S.}~\bibnamefont {Carretta}}, \bibinfo {author}
  {\bibfnamefont {P.}~\bibnamefont {Santini}}, \bibinfo {author} {\bibfnamefont
  {A.}~\bibnamefont {Stunault}}, \bibinfo {author} {\bibfnamefont
  {R.}~\bibnamefont {Caciuffo}}, \bibinfo {author} {\bibfnamefont
  {J.}~\bibnamefont {van Slageren}}, \bibinfo {author} {\bibfnamefont
  {B.}~\bibnamefont {Klemke}}, \bibinfo {author} {\bibfnamefont
  {A.}~\bibnamefont {Cousson}}, \bibinfo {author} {\bibfnamefont {G.~A.}\
  \bibnamefont {Timco}},\ and\ \bibinfo {author} {\bibfnamefont {R.~E.~P.}\
  \bibnamefont {Winpenny}},\ }\bibfield  {title} {\bibinfo {title} {Direct
  observation of finite size effects in chains of antiferromagnetically coupled
  spins},\ }\href {https://doi.org/10.1038/ncomms8061} {\bibfield  {journal}
  {\bibinfo  {journal} {Nature Communications}\ }\textbf {\bibinfo {volume}
  {6}},\ \bibinfo {pages} {7061} (\bibinfo {year} {2015})}\BibitemShut
  {NoStop}%
\bibitem [{\citenamefont {Garlatti}\ \emph {et~al.}(2020)\citenamefont
  {Garlatti}, \citenamefont {Allodi}, \citenamefont {Bordignon}, \citenamefont
  {Bordonali}, \citenamefont {Timco}, \citenamefont {Winpenny}, \citenamefont
  {Lascialfari}, \citenamefont {Renzi},\ and\ \citenamefont
  {Carretta}}]{Garlatti20}%
  \BibitemOpen
  \bibfield  {author} {\bibinfo {author} {\bibfnamefont {E.}~\bibnamefont
  {Garlatti}}, \bibinfo {author} {\bibfnamefont {G.}~\bibnamefont {Allodi}},
  \bibinfo {author} {\bibfnamefont {S.}~\bibnamefont {Bordignon}}, \bibinfo
  {author} {\bibfnamefont {L.}~\bibnamefont {Bordonali}}, \bibinfo {author}
  {\bibfnamefont {G.~A.}\ \bibnamefont {Timco}}, \bibinfo {author}
  {\bibfnamefont {R.~E.~P.}\ \bibnamefont {Winpenny}}, \bibinfo {author}
  {\bibfnamefont {A.}~\bibnamefont {Lascialfari}}, \bibinfo {author}
  {\bibfnamefont {R.~D.}\ \bibnamefont {Renzi}},\ and\ \bibinfo {author}
  {\bibfnamefont {S.}~\bibnamefont {Carretta}},\ }\bibfield  {title} {\bibinfo
  {title} {Breaking the ring: $^{53}${Cr}-{NMR} on the {Cr}$_8${Cd} molecular
  nanomagnet},\ }\href {https://doi.org/10.1088/1361-648x/ab7872} {\bibfield
  {journal} {\bibinfo  {journal} {Journal of Physics: Condensed Matter}\
  }\textbf {\bibinfo {volume} {32}},\ \bibinfo {pages} {244003} (\bibinfo
  {year} {2020})}\BibitemShut {NoStop}%
\bibitem [{\citenamefont {Kota}\ and\ \citenamefont {Imamura}(2016)}]{Kota16}%
  \BibitemOpen
  \bibfield  {author} {\bibinfo {author} {\bibfnamefont {Y.}~\bibnamefont
  {Kota}}\ and\ \bibinfo {author} {\bibfnamefont {H.}~\bibnamefont {Imamura}},\
  }\bibfield  {title} {\bibinfo {title} {Narrowing of antiferromagnetic domain
  wall in corundum-type {Cr}$_2${O}$_3$ by lattice strain},\ }\href
  {https://doi.org/10.7567/apex.10.013002} {\bibfield  {journal} {\bibinfo
  {journal} {Applied Physics Express}\ }\textbf {\bibinfo {volume} {10}},\
  \bibinfo {pages} {013002} (\bibinfo {year} {2016})}\BibitemShut {NoStop}%
\bibitem [{\citenamefont {Hirjibehedin}\ \emph {et~al.}(2007)\citenamefont
  {Hirjibehedin}, \citenamefont {Lin}, \citenamefont {Otte}, \citenamefont
  {Ternes}, \citenamefont {Lutz}, \citenamefont {Jones},\ and\ \citenamefont
  {Heinrich}}]{Hirjibehedin07}%
  \BibitemOpen
  \bibfield  {author} {\bibinfo {author} {\bibfnamefont {C.~F.}\ \bibnamefont
  {Hirjibehedin}}, \bibinfo {author} {\bibfnamefont {C.-Y.}\ \bibnamefont
  {Lin}}, \bibinfo {author} {\bibfnamefont {A.~F.}\ \bibnamefont {Otte}},
  \bibinfo {author} {\bibfnamefont {M.}~\bibnamefont {Ternes}}, \bibinfo
  {author} {\bibfnamefont {C.~P.}\ \bibnamefont {Lutz}}, \bibinfo {author}
  {\bibfnamefont {B.~A.}\ \bibnamefont {Jones}},\ and\ \bibinfo {author}
  {\bibfnamefont {A.~J.}\ \bibnamefont {Heinrich}},\ }\bibfield  {title}
  {\bibinfo {title} {Large magnetic anisotropy of a single atomic spin embedded
  in a surface molecular network},\ }\href
  {https://doi.org/10.1126/science.1146110} {\bibfield  {journal} {\bibinfo
  {journal} {Science}\ }\textbf {\bibinfo {volume} {317}},\ \bibinfo {pages}
  {1199} (\bibinfo {year} {2007})},\ \Eprint
  {https://arxiv.org/abs/https://www.science.org/doi/pdf/10.1126/science.1146110}
  {https://www.science.org/doi/pdf/10.1126/science.1146110} \BibitemShut
  {NoStop}%
\bibitem [{\citenamefont {Ueltzh{\"o}ffer}\ \emph {et~al.}(2016)\citenamefont
  {Ueltzh{\"o}ffer}, \citenamefont {Streubel}, \citenamefont {Koch},
  \citenamefont {Holzinger}, \citenamefont {Makarov}, \citenamefont {Schmidt},\
  and\ \citenamefont {Ehresmann}}]{Ueltzhoeffer16}%
  \BibitemOpen
  \bibfield  {author} {\bibinfo {author} {\bibfnamefont {T.}~\bibnamefont
  {Ueltzh{\"o}ffer}}, \bibinfo {author} {\bibfnamefont {R.}~\bibnamefont
  {Streubel}}, \bibinfo {author} {\bibfnamefont {I.}~\bibnamefont {Koch}},
  \bibinfo {author} {\bibfnamefont {D.}~\bibnamefont {Holzinger}}, \bibinfo
  {author} {\bibfnamefont {D.}~\bibnamefont {Makarov}}, \bibinfo {author}
  {\bibfnamefont {O.~G.}\ \bibnamefont {Schmidt}},\ and\ \bibinfo {author}
  {\bibfnamefont {A.}~\bibnamefont {Ehresmann}},\ }\bibfield  {title} {\bibinfo
  {title} {Magnetically patterned rolled-up exchange bias tubes: A paternoster
  for superparamagnetic beads},\ }\href
  {https://doi.org/10.1021/acsnano.6b03566} {\bibfield  {journal} {\bibinfo
  {journal} {{ACS} Nano}\ }\textbf {\bibinfo {volume} {10}},\ \bibinfo {pages}
  {8491} (\bibinfo {year} {2016})}\BibitemShut {NoStop}%
\bibitem [{hzd()}]{hzdr}%
  \BibitemOpen
  \href {http://www.hzdr.de} {\bibinfo {title} {{High Performance Computing at
  Helmholtz-Zentrum Dresden-Rossendorf}}}\BibitemShut {NoStop}%
\bibitem [{SLa()}]{SLaSi}%
  \BibitemOpen
  \href {http://slasi.knu.ua} {\bibinfo {title} {\textsf{SLaSi} spin--lattice
  simulations package}}\BibitemShut {NoStop}%
\end{thebibliography}
%

\end{document}